\documentclass{emulateapj}

\begin{document}

\newcommand{\mirlum}{L_{\rm 8}}
\newcommand{\ebmv}{E(B-V)}
\newcommand{\lha}{L(H\alpha)}
\newcommand{\lir}{L_{\rm IR}}
\newcommand{\lbol}{L_{\rm bol}}
\newcommand{\luv}{L_{\rm UV}}
\newcommand{\rs}{{\cal R}}
\newcommand{\ugr}{U_{\rm n}G\rs}
\newcommand{\ks}{K_{\rm s}}
\newcommand{\gmr}{G-\rs}
\newcommand{\wlya}{W_{\rm Ly\alpha}}
\newcommand{\lya}{\rm Ly\alpha}
\newcommand{\fhp}{f_{\rm 100}}
\newcommand{\fsp}{f_{\rm 160}}

\title{GOODS-{\em Herschel} Measurements of the Dust Attenuation of
  Typical Star-Forming Galaxies at High Redshift: Observations of
  UV-Selected Galaxies at $\lowercase{z}\sim 2$\altaffilmark{1}}

\author{\sc N. Reddy\altaffilmark{2,3}, M. Dickinson\altaffilmark{2},
D. Elbaz\altaffilmark{4}, G. Morrison\altaffilmark{5,6}, M. Giavalisco\altaffilmark{7},
R. Ivison\altaffilmark{8,9}, C. Papovich\altaffilmark{10}, D. Scott\altaffilmark{11}
V. Buat\altaffilmark{12}, D. Burgarella\altaffilmark{12}, V. Charmandaris\altaffilmark{13}, 
E. Daddi\altaffilmark{4}, 
G. Magdis\altaffilmark{4}, E. Murphy\altaffilmark{14}, B.~Altieri\altaffilmark{15}, 
H.~Aussel\altaffilmark{4}, H.~Dannerbauer\altaffilmark{4}, K.~Dasyra\altaffilmark{4},
H.S.~Hwang\altaffilmark{4}, J.~Kartaltepe\altaffilmark{2}, R.~Leiton\altaffilmark{4},
B.~Magnelli\altaffilmark{4}, P.~Popesso\altaffilmark{16}}

\altaffiltext{1}{{\em Herschel} is an ESA space observatory with
  science instruments provided by European-led Principal Investigator
  consortia and with important participation from NASA.}

\altaffiltext{2}{National Optical Astronomy Observatory, 950 N Cherry
  Ave, Tucson, AZ 85719, USA.}

\altaffiltext{3}{Hubble Fellow.}

\altaffiltext{4}{Laboratoire AIM-Paris-Saclay, CEA/DSM/Irfu - CNRS -
  Universit\'{e} Paris Diderot, CE-Saclay, F-91191, Gif-sur-Yvette,
  France.}
\altaffiltext{5}{Institute for Astronomy, University of Hawaii, Honolulu, HI 96822, USA.}
\altaffiltext{6}{Canada-France-Hawaii Telescope, Kamuela, HI, 96743, USA.}
\altaffiltext{7}{University of Massachusetts, Amherst, Astronomy Department, Amherst, MA 01003, USA.}
\altaffiltext{8}{UK Astronomy Technology Centre, Science and Technology Facilities Council, Royal
Observatory, Blackford Hill, Edinburgh EH9 3HJ, UK.}
\altaffiltext{9}{Institute for Astronomy, University of Edinburgh, Blackford Hill, Edinburgh EH9 3HJ,
UK.}
\altaffiltext{10}{Texas A\&M University, Department of Physics and Astronomy, College
Station, TX 77845, USA.}
\altaffiltext{11}{Department of Physics and Astronomy, University of British Columbia, Vancouver,
BC V6T 1Z1, Canada.}
\altaffiltext{12}{Laboratoire d'Astrophysique de Marseille, OAMP, Universit\'{e} Aix-Marseille, CNRS,
38 Rue Fr\'{e}d\'{e}ric Joliot-Curie, 13388 Marseille Cedex 13, France.}
\altaffiltext{13}{University of Crete, Department of Physics, GR-71003, Heraklion, Greece.}
\altaffiltext{14}{{\em Spitzer} Science Center, MC 314-6, California Institute of Technology,
Pasadena, CA 91125.}
\altaffiltext{15}{{\em Herschel} Science Centre, European Space Astronomy Centre, Villanueva de
la Ca\~{n}ada, 28691 Madrid, Spain.}
\altaffiltext{16}{Max-Planck-Institute f\"{u}r Extraterrestrische Physik (MPE), Postfach 1312, 85741,
Garching, Germany.}

\slugcomment{DRAFT: \today}

\begin{abstract}

We take advantage of the sensitivity and resolution of the {\em
  Herschel Space Observatory} at $100$ and $160$\,$\mu$m to directly
image the thermal dust emission and investigate the infrared
luminosities ($\lir$) and dust obscuration of typical star-forming
($L^{\ast}$) galaxies at high redshift.  Our sample consists of $146$
UV-selected galaxies with spectroscopic redshifts $1.5\le z_{\rm
  spec}<2.6$ in the GOODS-North field.  Supplemented with deep Very
Large Array (VLA) and {\em Spitzer} imaging, we construct median
stacks at the positions of these galaxies at $24$, $100$, and
$160$\,$\mu$m, and $1.4$\,GHz.  The comparison between these stacked
fluxes and a variety of dust templates and calibrations implies that
typical star-forming galaxies with UV luminosities $\luv \ga
10^{10}$\,L$_{\odot}$ at $z\sim 2$ are luminous infrared galaxies
(LIRGs) with a median $\lir= (2.2\pm 0.3)\times 10^{11}$\,L$_{\odot}$.
Their median ratio of $\lir$ to rest-frame $8$\,$\mu$m luminosity
($\mirlum$) is $\lir/\mirlum = 8.9\pm1.3$ and is $\approx 80\%$ larger
than that found for most star-forming galaxies at $z\la 2$.  This
apparent redshift evolution in the $\lir/\mirlum$ ratio may be tied to
the trend of larger infrared luminosity surface density for $z\ga 2$
galaxies relative to those at lower redshift.  Typical galaxies at
$1.5\le z<2.6$ have a median dust obscuration $\lir/\luv = 7.1\pm1.1$,
which corresponds to a dust correction factor, required to recover the
bolometric star formation rate (SFR) from the unobscured UV SFR, of
$5.2\pm0.6$.  This result is similar to that inferred from previous
investigations of the UV, H$\alpha$, $24$\,$\mu$m, radio, and X-ray
properties of the same galaxies studied here.  Stacking in bins of UV
slope ($\beta$) implies that $L^{\ast}$ galaxies with redder spectral
slopes are also dustier, and that the correlation between $\beta$ and
dustiness is similar to that found for local starburst galaxies.
Hence, the rest-frame $\simeq 30$ and $50$\,$\mu$m fluxes validate on
average the use of the local UV attenuation curve to recover the dust
attenuation of typical star-forming galaxies at high redshift.  In the
simplest interpretation, the agreement between the local and high
redshift UV attenuation curves suggests a similarity in the dust
production and stellar and dust geometries of starburst galaxies over
the last $10$\,billion years.

\end{abstract}


\section{INTRODUCTION}
\label{sec:intro}

A key aspect of measuring total star formation rates is to assess how
a galaxy's spectral energy distribution is modulated by the effects of
interstellar dust.  The ability to quantify dust attenuation at high
redshift is hindered by the limited dynamic range in luminosity at
wavelengths that are sensitive to dust emission.  While the Lyman
Break technique \citep{steidel95} has proven to be the most effective
method of identifying galaxies across a large dynamic range in
luminosity and lookback time, it requires observations in the
rest-frame UV: the massive stars giving rise to the UV continuum also
produce much of the dust that attenuates this continuum.  The
diminution of UV light is compounded by the fact that much of this
radiation is re-emitted in the far-infrared where current instrumental
sensitivity is insufficient to directly detect typical star-forming
galaxies at $z\ga 2$.  It has therefore become common to use local
relations between monochromatic and bolometric luminosity to infer the
star formation rates and dust attenuation of high-redshift galaxies.

Prior to the era of large-scale multi-wavelength surveys like GOODS
\citep{dickinsongoods03,giavalisco04}, inferring the dust attenuation
at high redshift commonly entailed using relations between the
variation of the UV continuum slope with dustiness as found for local
starburst galaxies \citep{meurer99, calzetti00, heckman98}.  However,
such correlations were previously untested at high redshift.  The
correlation between the redness of the UV slope ($\beta$) and dust
attenuation has limited applicability when examined over a larger
range in galaxy stellar population and luminosity.  Galaxies with
older and less massive stars contributing significantly to the UV
emission can also exhibit a redder spectral slope \citep{calzetti97,
  kong04, buat05}.  Further, the UV slope decouples from extinction
for very luminous galaxies where virtually all of the star formation
is obscured, such as is the case for low redshift ultra-luminous
infrared galaxies (ULIRGs; \citealt{goldader02}).  Nonetheless, the
local trend between UV slope and dustiness is still applied widely to
galaxies at high redshift; this correlation is often the only means by
which one can infer the dust attenuation of galaxies at $z\ga 3$,
where the dust emission from a typical galaxy may be several orders of
magnitude below the sensitivity limits of current infrared
instruments.

Incremental progress in determining dust attenuation at high redshift
was achieved with the first ultra-deep radio and X-ray surveys (e.g.,
\citealt{richards00, alexander03}).  Though the sensitivity (even at
GOODS-depth) at these wavelengths was insufficient to individually
detect typical star-forming galaxies at $z\ga 2$, these surveys did
allow for estimates of the mean dust attenuation based on stacking
analyses.  Initial studies suggested a general agreement between UV
and radio/X-ray inferences of dust attenuation at $z\sim 2-3$
\citep{nandra02, seibert02, reddy04, reddy05a, reddy06a, daddi07a,
  pannella09}.

The launch of the {\em Spitzer Space Telescope} enabled the first
direct detection of the dust emission in non-lensed $L^{\ast}$
galaxies at $z\sim 2$; at these redshifts, {\em Spitzer}'s MIPS
$24$\,$\mu$m band is sensitive to the strongest dust emission feature
(at $7.7$\,$\mu$m) in star-forming galaxies.  This feature arises from
the stochastic UV photo-heating of small dust grains and hydrocarbons
(e.g., \citealt{puget89, tielens99}) and is found to correlate with
the UV radiation from OB stars (e.g., \citealt{forster04, roussel01}),
albeit with significant scatter (e.g., \citealt{kennicutt09, hogg05,
  helou01, engelbracht05, normand95}).  Several studies have
demonstrated that the dust attenuation inferred from the UV spectral
slope, $\beta$, for typical $z\sim 2$ galaxies is in general agreement
with that inferred from MIPS $24$\,$\mu$m (e.g., \citealt{reddy06a,
  reddy10a, daddi07a}).  Regardless, local studies of resolved
galaxies have emphasized the complexity of the $8$\,$\mu$m emission
and the need to combine it with other obscured (IR) and unobscured
(UV, H$\alpha$) tracers of star formation in order to obtain the most
reliable measure of the bolometric luminosity \citep{kennicutt09}.

To this end, we take advantage of the improved sensitivity of {\em
  Herschel}/PACS at $100$ and $160$\,$\mu$m \citep{pilbratt10,
  poglitsch10} to measure directly for the first time the thermal dust
emission from a large sample of typical ($L^{\ast}$) star-forming
galaxies at $z\sim 2$.  We make use of the deep PACS data made
possible by the GOODS-{\em Herschel} Open Time Key project (PI:
D. Elbaz).  We supplement these data with deep Very Large Array (VLA)
$1.4$\,GHz continuum imaging in the GOODS-North field
\citep{morrison10}.  The primary aim is to constrain the average
infrared luminosities of galaxies at high redshift, particularly those
selected by their rest-frame UV colors, and to compare them to
UV-based inferences.  We begin in Section~\ref{sec:sample} by
discussing the UV-selected galaxies and previous efforts to measure
their stellar populations and dust attenuation.  We also provide a
brief description of the {\em Herschel} and radio data.  Our stacking
method and stacking simulations are described in
Section~\ref{sec:stacking}, followed by a discussion of the dust
spectral energy distribution (SED) fits to the stacked fluxes and the
total infrared luminosities in Section~\ref{sec:luminosities}.  In
Section~\ref{sec:discussion} we proceed to compare the {\em
  Herschel}-based inferences of the dust attenuation with that
inferred from the UV slope and discuss systematics in the dust
obscuration with stellar population age and bolometric and UV
luminosity.  For ease of comparison with other studies, we assume a
\citet{salpeter55} initial mass function (IMF) and adopt a cosmology
with $H_{0}=70$\,km\,s$^{-1}$\,Mpc$^{-1}$, $\Omega_{\Lambda}=0.7$, and
$\Omega_{\rm m}=0.3$.

\section{SAMPLE SELECTION AND {\em Herschel} PACS 100 and 160\,$\mu$m Data}
\label{sec:sample}

\subsection{Rest-Frame UV-Selected Sample of Star-Forming Galaxies at $z\sim 2-3$}

The GOODS-North field was targeted as part of an ongoing imaging and
spectroscopic survey of UV-selected galaxies at $z\sim 2-3$
\citep{steidel04}, primarily to take advantage of the extensive
multi-wavelength data that exist for this field.  Spectroscopic
catalogs and analysis of the GOODS-North sample, including the stellar
masses and dust attenuation of $L^{\ast}$ galaxies at $z\sim 2-3$, are
presented in \citet{reddy06b}.  Briefly, the galaxies were selected to
lie at redshifts $1.5\la z\la 3.4$ based on the ``BM,'' ``BX,'' and
Lyman Break galaxy (LBG) color criteria \citep{steidel03, steidel04,
  adelberger04}.  Extensive spectroscopic followup of candidates
brighter than $\rs = 25.5$ was conducted using the blue arm of the Low
Resolution Imaging Spectrograph (LRIS; \citealt{steidel04}) on the
Keck I telescope.

The criteria used to construct our sample are identical to those of
\citet{reddy10a}.  Only galaxies with secure spectroscopic redshifts
$1.5\le z_{\rm spec}\le 2.6$ were considered as this is the redshift
range over which the MIPS band is sensitive to rest-frame $8$\,$\mu$m
emission.  Galaxies with any AGN signatures based on optical emission
lines (Ly$\alpha$, CIV, NV) or a power-law SED through the IRAC and
MIPS $24$\,$\mu$m bands were excluded.  Additionally, galaxies were
included only if we were able to obtain robust point spread function
(PSF) fits to their $24$\,$\mu$m emission; in practice, this meant
that a few galaxies were rejected because they are confused with
nearby neighbors (see \citealt{reddy10a} for the parameter used to
measure the degeneracy of the PSF fits).  This sample forms the basis
of our stacking analysis, and includes a total of 146 galaxies,
approximately $30\%$ of which are detected with greater than
$3$\,$\sigma$ significance in the GOODS-N $24$\,$\mu$m data.

\subsection{Properties of the Sample and SED Modeling}

\citet{reddy09} use a larger spectroscopic and photometric sample (the
latter extending to $\rs\sim 27.0$) in 31 independent fields to
quantify the selection function for the UV sample and determine the UV
luminosity function at $z\sim 2$.  This analysis indicates that
UV-selected galaxies commonly targeted for spectroscopy with $\rs\la
25.5$ have luminosities similar to $L^{\ast}$ of the UV luminosity
function, where $L^{\ast}_{\rm UV}\approx 4\times
10^{10}$\,L$_{\odot}$ \citep{reddy09}.  Throughout the text,
$L^{\ast}$ refers to the characteristic luminosity of the UV
luminosity function, unless stated otherwise.  The stellar populations
of these galaxies have been modeled previously, and we refer the
reader to \citet{reddy10a} for a detailed description of the modeling
procedure.

Briefly, Charlot \& Bruzual (in prep.) stellar population models with
a range of exponentially-declining star formation histories $\tau =$
10, 20, 50, 100, 200, 500, 1000, 2000, and 5000\,Myr, as well as a
constant star formation history, were fit to the observed $U_{\rm
  n}G{\cal R}$, $J\ks$, and {\em Spitzer}/IRAC data.  In addition, we
considered ages spaced roughly logarithmically between 70 and
5000\,Myr, excluding any that are older than the age of the Universe
at the redshift of each galaxy.  The lower limit to the allowed ages
($70$\,Myr) was adopted to reflect the typical dynamical time-scale as
inferred from velocity dispersion and size measurements of $z\sim 2$
LBGs \citep{erb06c, law07}.  Reddening was incorporated by employing
the \citet{calzetti00} attenuation curve and allowing $\ebmv$ to range
between 0.0 and 0.6.

We adopt the best-fit stellar population parameters obtained when
assuming a constant star formation (CSF) history, unless an
exponentially declining ($\tau$) model gives a significantly better
fit to the broadband data.  Generally, the $\chi^2$ values assuming
the CSF model were similar to those obtained when $\tau$ is allowed to
vary freely.  Further, more extreme star formation histories, where
the ratio of the age to $\tau$ is much greater than unity, i.e.,
$t_{\rm age}/\tau \gg 1$, can be ruled out based on the presence of O
star and Wolf-Rayet features in the composite UV spectra of $z\sim 2$
galaxies, as well as the fact that such models often predict ages that
are unrealistically much younger than the dynamical timescale at
$z\sim 2$.  For the present analysis, we use the results of the
stellar population modeling to distinguish those galaxies that have
young star formation ages $\la 100$\,Myr.  Such galaxies have been
shown previously to depart from the UV attenuation curve established
for more typical galaxies at $z\sim 2$ \citep{reddy10a, reddy06a}, and
we wish to explore this difference with the {\em Herschel} data.

\subsection{Construction of Subsamples}

The sample of 146 UV-selected galaxies was subdivided into subsamples
in order to investigate the differences in dust attenuation between
galaxies with blue and red UV spectral slopes, those with high
bolometric luminosities ($\lbol \equiv \lir + \luv$), and those with
young stellar population ages.  The properties of these subsamples,
including the criteria used to construct them, the number of galaxies
in each subsample, and their $\beta$ and redshift distributions, are
summarized in Table~\ref{tab:stacks}.  As noted above, the ages are
estimated from stellar population modeling of the rest-frame UV
through {\em Spitzer} IRAC photometry.  For the purpose of contructing
the subsamples, we estimated bolometric luminosities based on the
\citet{reddy10a} calibration between $8$\,$\mu$m and infrared
luminosity (see next section).  Finally, UV slopes $\beta$ were
determined from the $G-{\cal R}$ colors of galaxies as follows.  We
generated power laws in $f(\lambda)\propto \lambda^\beta$ for $-2.5\le
\beta\le 1.0$ with $\Delta\beta = 0.01$.  These were attenuated for
the Ly$\alpha$ forest opacity assuming the \citet{madau95}
prescription and multiplied by the $G$ and ${\cal R}$ transmission
filters.  The $G$ band is affected by the Ly$\alpha$ forest only for
redshifts $z=2.5$; statistical fluctuations in the forest will not
affect $\beta$ as most of the galaxies considered here lie below
redshift $z\le 2.5$.  For this same reason, Ly$\alpha$ lies outside of
the $G$-band filter and will therefore not affect $\beta$.  The UV
slope for a given galaxy is taken to be the one which gives the
closest match in $G-{\cal R}$ color to the observed value.  The error
in UV slope is related directly to the error in color and is typically
$\sigma_\beta\simeq 0.11$.

\begin{deluxetable*}{llll}
\tabletypesize{\scriptsize}
\tablewidth{0pc}
\tablecaption{Properties of the Stacks I: $100$\,$\mu$m Fluxes, Redshifts, and UV Slopes}
\tablehead{
\colhead{} &
\colhead{} &
\colhead{$\fhp$} &
\colhead{$\fhp$} \\
\colhead{Sample} & 
\colhead{Criterion~\tablenotemark{a}} &
\colhead{$r_{\rm exc}=3\farcs43$~\tablenotemark{b}} &
\colhead{No Exclusion~\tablenotemark{c}}}
\startdata
{\bf A.} &  All & $(3.0\pm0.5)\times 10^{-4}$\,Jy & $(3.0\pm0.4)\times 10^{-4}$\,Jy \\
{\bf ALL} & & $N=106$ & $N=146$ \\
{\bf UV-} & & $\langle \beta\rangle=-1.41\pm0.42$ & $\langle \beta\rangle=-1.41\pm0.41$ \\
{\bf SELECTED} & & $\langle z\rangle = 2.08\pm0.24$ & $\langle z\rangle = 2.08\pm0.26$ \\
\\
{\bf B.} & Age $\ga 100$\,Myr; & $(2.4\pm0.5)\times 10^{-4}$\,Jy & $(2.8\pm0.5)\times 10^{-4}$\,Jy \\
{\bf $L^{\ast}$} & $L_{\rm bol}\le 10^{12}$\,L$_{\odot}$ & $N=83$ & $N=114$ \\
{\bf GALAXIES} & & $\langle \beta\rangle=-1.48\pm0.37$ & $\langle \beta\rangle=-1.46\pm0.38$ \\
& & $\langle z\rangle=2.09\pm0.24$ & $\langle z\rangle=2.09\pm0.25$ \\
\\
{\bf C.} & Age $\ga 100$\,Myr; & $(1.9\pm0.7)\times 10^{-4}$\,Jy & $(2.1\pm0.7)\times 10^{-4}$\,Jy \\
{\bf BLUE} & $L_{\rm bol}\le 10^{12}$\,L$_{\odot}$; & $N=51$ & $N=67$ \\
{\bf UV-SLOPES} & $\beta < -1.4$ & $\langle \beta\rangle = -1.61\pm0.21$ & $\langle \beta\rangle = -1.59\pm0.21$ \\
& & $\langle z\rangle = 2.05\pm0.24$ & $\langle z \rangle = 2.05\pm0.26$ \\
\\
{\bf D.} & Age $\ga 100$\,Myr; & $(3.8\pm0.9)\times 10^{-4}$\,Jy & $(3.4\pm0.6)\times 10^{-4}$\,Jy \\
{\bf RED} & $L_{\rm bol}\le 10^{12}$\,L$_{\odot}$; & $N=32$ & $N=47$ \\
{\bf UV-SLOPES} & $\beta \ge -1.4$ & $\langle \beta\rangle = -1.10\pm0.26$ & $\langle \beta\rangle = -1.15\pm0.26$ \\
& & $\langle z\rangle = 2.19\pm0.22$ & $\langle z \rangle = 2.14\pm0.24$ \\
\\
{\bf E.} & Age $\ga 100$\,Myr; & $(11.0\pm1.7)\times 10^{-4}$\,Jy & $(9.9\pm1.2)\times 10^{-4}$\,Jy \\
{\bf ULIRGs} & $L_{\rm bol}>10^{12}$\,L$_{\odot}$ & $N=9$ & $N=12$ \\
& & $\langle \beta\rangle=-0.84\pm0.48$ & $\langle\beta\rangle=-0.84\pm0.46$ \\
& & $\langle z\rangle=2.03\pm0.27$ & $\langle z\rangle=2.22\pm0.29$ \\
& & $\langle L_{\rm bol}\rangle=(1.3\pm0.1)\times 10^{12}$\,L$_{\odot}$ & $\langle L_{\rm bol}\rangle=(1.3\pm0.1)\times 10^{-4}$\,L$_{\odot}$ \\
& & $L_{\rm bol}^{\rm max} = 1.8\times 10^{12}$\,L$_{\odot}$ & $L_{\rm bol}^{\rm max} = 1.8\times 10^{12}$\,L$_{\odot}$ \\
\\
{\bf F.} & Age $\la 100$\,Myr & 3\,$\sigma$: $<4.4\times10^{-4}$\,Jy & 3\,$\sigma$: $<3.2\times10^{-4}$\,Jy \\
{\bf YOUNG} & & $N=14$ & $N=20$ \\
{\bf GALAXIES} & & $\langle \beta\rangle= -1.17\pm0.40$ & $\langle \beta\rangle= -1.30\pm0.40$ \\
& & $\langle z\rangle=2.00\pm0.26$ & $\langle z\rangle = 2.00\pm 0.28$ \\
\enddata
\tablenotetext{\,}{NOTE. -- Each entry includes (a) the stacked flux at $100$\,$\mu$m and 
its measurement uncertainty; (b) the number of galaxies contributing to the stack; and (c) the mean and sample dispersion 
of the UV slopes ($\beta$) and redshifts of those galaxies.  We assign 3\,$\sigma$ upper limits to fluxes 
in cases where the $1$\,$\sigma$ measurement
uncertainty is larger than the stacked flux.  For Sample E (ULIRGs), we include the mean and error 
in the mean of the bolometric luminosity ($\lbol$; derived using the calibration of \citealt{reddy10a}) 
and the maximum bolometric luminosity ($\lbol^{\rm max}$) contributing to the stack.}
\tablenotetext{a}{The criteria used to select each subsample are listed here.  The ages were determined
from SED-fitting to the broadband photometry (see text) and the bolometric luminosities, $\lbol$, were
determined from our previous calibration between $24$\,$\mu$m and infrared luminosity \citep{reddy10a}.}
\tablenotetext{b}{Criteria for excluding galaxies from the stack.  If any galaxy lies within a
distance $r_{\rm exc}$ of a nearby optical, $\ks$-band, or IRAC source, it is excluded from the stack.}
\tablenotetext{c}{No exclusion radius adopted; all galaxies are median stacked.}
\label{tab:stacks}
\end{deluxetable*}

\subsection{Previous MIPS Results}

To provide a context for our present analysis, we briefly summarize
previous efforts to constrain the dust emission and bolometric
luminosities of galaxies at $z\sim 2-3$.  \citet{reddy06a}
investigated the variation in $24$\,$\mu$m flux with dust-corrected UV
and X-ray measures of the bolometric luminosities of the same
UV-selected galaxies analyzed here.  Extending upon this result,
\citet{reddy10a} used a sample of 90 Lyman Break galaxies to examine
the relationship between rest-frame $8$\,$\mu$m ($\nu L_{\nu}[8\mu
  {\rm m}]\equiv \mirlum$) and H$\alpha$ luminosity at $z\sim 2$,
finding a tight trend between the two, with a scatter of $\approx
0.24$\,dex.  Using H$\alpha$ luminosity as a proxy for total star
formation rate -- after accounting for dust with a \citet{calzetti00}
attenuation curve -- then allowed these authors to establish a
relationship between $\mirlum$ and dust obscured SFR, or $\lir$.

The previous studies found that typical ($L^{\ast}$) star-forming
galaxies at $z\sim 2$ have dust attenuations, or infrared-to-UV
luminosity ratios, $\lir/\luv \simeq 5$, similar to the values
predicted from the UV slope, $\beta$, using the local correlation
between $\beta$ and $\lir/\luv$ \citep{meurer99}.  Investigations of
the $24$\,$\mu$m emission of galaxies selected at rest-frame optical
wavelengths -- resulting in samples that are not orthogonal to the one
analyzed here -- have reached similar conclusions regarding the
validity of UV-based dust corrections for moderately-luminous galaxies
at $z\sim 2$ \citep{daddi07a}.  These results have been extended to
higher redshift: \citet{magdis10a} demonstrate that UV-based dust
corrections for LBGs at $z\sim 3$ yield bolometric luminosities
comparable to those inferred from infrared, radio, and millimeter
measures.  The $160$~$\mu$m emission from $z\sim 3$ LBGs also
corresponds to $\lir$ that are similar to those obtained with
$24$\,$\mu$m estimates \citep{magdis10b}.  While this is important
validation of our understanding of star formation and dust obscuration
at high redshift, the LBGs that \citet{magdis10b} detected in stacked
$160$\,$\mu$m data have significantly higher luminosities ($\langle
\lir\rangle \approx 1.6\times 10^{12}$\,L$_{\odot}$) and star
formation rates ($> 100$\,M$_{\odot}$\,yr$^{-1}$) than those of the
typical, $L^{\ast}$, galaxies analyzed here.

In any case, while this agreement is encouraging, and instills
confidence in our ability to recover the dust attenuation of typical
galaxies from measurements of their UV continuum slopes, uncertainties
in the {\em k}-corrections and the conversion between $8$\,$\mu$m and
infrared luminosity (as well as conversion between X-ray emission and
SFR, for the X-ray stacking analyses) no doubt introduce some scatter
in the inferred dust obscurations.  The primary goal of our present
analysis to obtain more direct measures of the total infrared
luminosities of $z\sim 2$ galaxies using {\em Herschel} data, which
are described below.

\subsection{{\em Herschel} and VLA Data}

The GOODS-{\em Herschel} Open Time Key Program (PI: D. Elbaz) includes
$\approx 124$\,hrs of $100$ and $160$\,$\mu$m PACS imaging and $31$\,hrs
of SPIRE imaging in the GOODS-North field.  The $3$\,$\sigma$ depths of
the PACS 100 and $160$\,$\mu$m images are 1.1\,mJy and 2.6\,mJy,
respectively.  None of the galaxies in the sample analyzed here are
detected to these depths.  Further details on the data reduction are
given in \citet{elbaz11}.  PSFs were constructed based on the catalogs
of {\em Herschel} detections discussed in \citet{elbaz11}.  When
performing PSF photometry, we adjusted all fluxes upward by
multiplicative factors of 1.37 and 1.29 at $100$ and $160$\,$\mu$m
respectively, to account for missing flux in the wings of the PSFs.
The similar factor for the MIPS $24$\,$\mu$m PSF is 1.22.

The VLA radio 1.4\,GHz data are described in \citet{morrison10}.
Briefly, a total of $165$\,hrs of observations, including $42$\,hr
from \citet{richards00}, were combined to produce a radio map of the
GOODS-North field.  The noise level of the image is $\sim
3.9$\,$\mu$Jy\,beam$^{-1}$ near the center and $\sim
8$\,$\mu$Jy\,beam$^{-1}$ at 15$\arcmin$ from the center.  The
synthesized beamsize of the radio data is $\sim 1\farcs7 \times
1\farcs6$, corresponding to a physical scale of $14.1\times 13.3$\,kpc
at $z=2.30$.  The radio map used in the stacking analysis is primary
beam corrected.

\begin{figure}
\plotone{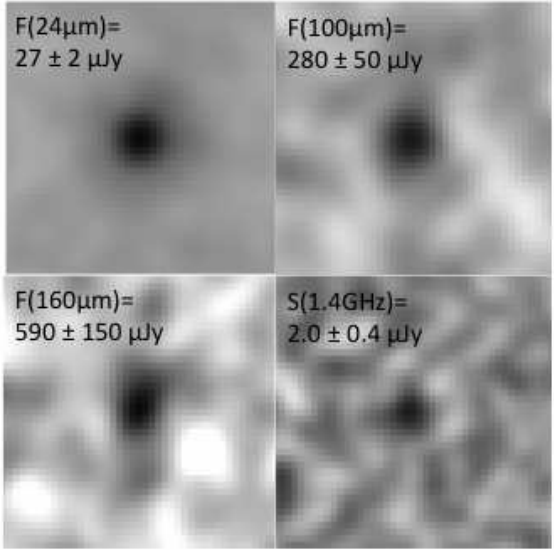}
\caption{PSF-convolved median stacked images for ``Sample B''
  ($L^{\ast}$ galaxies) at $24$, $100$, $160$\,$\mu$m, and $1.4$\,GHz,
  with pixel scales of $1\farcs2$, $1\farcs2$, $2\farcs4$, and
  $0\farcs5$, respectively.}
\label{fig:stackpost}
\end{figure}

\section{Stacking Method and Simulations}
\label{sec:stacking}

\subsection{General Stacking Procedure}

The method used to stack the {\em Herschel} $100$ and $160$\,$\mu$m,
{\em Spitzer} $24$\,$\mu$m, and the VLA radio data is as follows.  We
extracted an area $50\arcsec \times 50\arcsec$ around each object to
be stacked.  The area was chosen to be large enough to obtain a
reliable estimate of the local background.  These cutouts were then
median combined after rotating each sub-image by $90\degr$ relative
the previous sub-image contributing to the stack, in order to minimize
image defects that are aligned with the scanning orientation of the
data acquisition.\footnote{To test for any systematic effect
  associated with asymmetries in the PSF, we also stacked without
  rotating each sub-image.  The results obtained with and without
  rotating the sub-images were identical within the uncertainties of
  the stacked flux measurements.}  We adopted the median stacks to
ensure the results are not biased by bright outliers.  We stacked the
$24$\,$\mu$m data in exactly the same way as was done for the $100$
and $160$\,$\mu$m data, irrespective of whether a galaxy was directly
detected at $24$\,$\mu$m, in order to ensure the most consistent
results across wavelengths.\footnote{For the most conservative
  estimate of the stacked radio flux, we did not correct for bandwidth
  smearing (e.g., \citealt{pannella09}), rather we fit an elliptical
  Gaussian to the stacked light profile and adopted the peak value for
  the radio flux.}

To test the effects of a non-negligible amount of flux in the wings of
the stacked signal relative to the PSF (due to positional
uncertainty), we performed both PSF-fitting photometry and simple
aperture photometry on the stacked flux.  The background and
measurement uncertainty in the stacked image are taken respectively to
be the mean and $1$\,$\sigma$ dispersion in flux computed by fitting
many PSFs at random positions in the stacked image.  We simulatenously
fit the target flux (with a PSF) and a constant value to account for
the background level in the stacked image.  We found no significant
systematic bias in the aperture versus PSF-derived flux relative to
the measurement uncertainties, and we adopted the PSF-determined flux
measurements.  For the radio data, the PSF-measured flux and total
integrated flux differ by $\approx 20-30\%$ due to bandwidth smearing.
To account for this effect, we have adopted the fluxes computed from
fitting the best-fit elliptical Gaussian profile (with peak intensity
normalized to unity) to the stacked radio emission, yielding the
integrated flux density.

\subsection{Residual Images}

Residual maps at $100$ and $160$\,$\mu$m were constructed in order to
ensure the most robust determinations of the stacked
fluxes.\footnote{We did not construct a residual map for the VLA data
  given the higher resolution of these data and the lower $1.4$\,GHz
  source surface density.}  Sources detected at $>3$\,$\sigma$ were
subtracted from the science mosaics (none of the galaxies in our
sample are detected at $100$ and $160$\,$\mu$m at this level).  We
considered detections based both on a blind catalog
(``blind-subtracted''), where a detection algorithm was used directly
on the PACS mosaics, as well as a catalog constructed using
$24$\,$\mu$m priors to define the positions of sources in the longer
wavelength PACS bands (``prior-substracted'').  In both cases, we used
PSF photometry to determine the fluxes of sources and subtract them
from the science images.  To ensure that there are no systematic
effects that may bias the flux measurement made on the residual maps,
we performed the same stacking analysis on the science images
themselves.  Comparison between the stacks on the blind-residual maps,
prior-residual maps, and the science images, showed that the stacked
fluxes were within $10\%$ of each other.  Typically, the stacked
fluxes measured on the science images were {\em lower} than those
measured on the residual images.  This effect is attributed to the
higher background level in the stack derived from the science images
due to the PSF wings of nearby sources.  The similarity in flux
regardless of the image used for the stack suggests that the median
combination of sub-images is robust to bright outliers, as expected.
The flux measurements obtained by stacking on the prior-fit residual
maps are adopted for further analysis.  The median stacked data for
$L^{\ast}$ galaxies (Sample B; Table~\ref{tab:stacks}) at $24$, $100$,
and $160$\,$\mu$m, and $1.4$\,GHz, are shown in
Figure~\ref{fig:stackpost}.

\subsection{Exclusion Tests and Simulations}

A common concern in stacking data with relatively poor resolution is
the potential contribution from unrelated sources within the beam that
are formally undetected (i.e., with a $<3\,\sigma$ significance) and
which therefore remain in the residual images, but which may
contribute significantly to the stacked flux.  If these unrelated
sources have a random spatial distribution, then their contribution to
the stacked flux will be accounted for when we subtract the background
in the stacked images.  However, if the unrelated sources are
clustered with respect to the UV-selected galaxies, they will
contribute to the stacked fluxes.  To test for the effect of
clustering on the stacked fluxes, we considered a number of tests, as
we describe below.

\begin{deluxetable*}{lrrrr}
\tabletypesize{\scriptsize}
\tablewidth{0pc}
\tablecaption{Properties of the Stacks II: Summary of Mid-IR, IR, and Radio Fluxes}
\tablehead{
\colhead{} &
\colhead{$f_{\rm 24}$} &
\colhead{$f_{\rm 100}$} &
\colhead{$f_{\rm 160}$} &
\colhead{$f_{\rm 1.4}$} \\
\colhead{Sample} &
\colhead{(Jy)} &
\colhead{(Jy)} &
\colhead{(Jy)} &
\colhead{(Jy)}}
\startdata
{\bf A.} & $(33\pm4)\times 10^{-6}$ & $(3.0\pm0.4)\times 10^{-4}$ & $(7.3\pm1.2)\times 10^{-4}$ & $(2.1\pm0.5)\times 10^{-6}$ \\
{\bf B.} & $(27\pm2)\times 10^{-6}$ & $(2.8\pm0.5)\times 10^{-4}$ & $(5.9\pm1.5)\times 10^{-4}$ & $(2.0\pm0.4)\times 10^{-6}$ \\
{\bf C.} & $(21\pm3)\times 10^{-6}$ & $(2.1\pm0.7)\times 10^{-4}$ & $(5.4\pm1.6)\times 10^{-4}$ & $(1.6\pm0.6)\times 10^{-6}$ \\
{\bf D.} & $(35\pm3)\times 10^{-6}$ & $(3.4\pm0.6)\times 10^{-4}$ & $(6.6\pm2.1)\times 10^{-4}$ & $(2.6\pm0.7)\times 10^{-6}$ \\
{\bf E.} & $(130\pm4)\times 10^{-6}$ & $(9.9\pm1.2)\times 10^{-4}$ & $(22.4\pm3.8)\times 10^{-4}$ & $(6.6\pm1.3)\times 10^{-6}$ \\
{\bf F.} & $(10\pm2)\times 10^{-6}$ & 3\,$\sigma$: $<3.2\times 10^{-4}$ & $(4.6\pm3.3)\times 10^{-4}$ & 3$\,\sigma$: $<3.1\times 10^{-6}$ \\
\enddata
\tablenotetext{\,}{NOTE. -- The quoted errors reflect measurement uncertainty in the stacked fluxes.}
\label{tab:fluxes}
\end{deluxetable*}

\begin{deluxetable*}{lcccc}
\tabletypesize{\scriptsize}
\tablewidth{0pc}
\tablecaption{Properties of the Stacks III: Biases and Errors in Stacked Fluxes}
\tablehead{
\colhead{} &
\colhead{$24$\,$\mu$m} &
\colhead{$100$\,$\mu$m} &
\colhead{$160$\,$\mu$m} &
\colhead{$1.4$\,GHz} \\
\colhead{Sample} &
\colhead{1-Bias~\tablenotemark{a} / Error~\tablenotemark{b}} &
\colhead{1-Bias~\tablenotemark{a} / Error~\tablenotemark{b}} &
\colhead{1-Bias~\tablenotemark{a} / Error~\tablenotemark{b}} &
\colhead{1-Bias~\tablenotemark{a} / Error~\tablenotemark{b}}}
\startdata
{\bf A.} & 0.94 / 0.002 & 0.96 / 0.012 & 0.92 / 0.014 & 1.03 / 0.016 \\
{\bf B.} & 0.94 / 0.003 & 0.95 / 0.017 & 0.92 / 0.022 & 1.02 / 0.024 \\
{\bf C.} & 0.94 / 0.015 & 0.96 / 0.039 & 0.92 / 0.040 & 1.03 / 0.054 \\
{\bf D.} & 0.94 / 0.004 & 0.96 / 0.035 & 0.92 / 0.045 & 1.03 / 0.038 \\
{\bf E.} & 0.95 / 0.009 & 0.95 / 0.046 & 0.93 / 0.058 & 1.05 / 0.075 \\
\enddata
\tablenotetext{a}{Average bias of stacked flux, defined as the ratio of the mean measured flux
to the simulated flux, $\langle f^{meas}\rangle / f^{sim}$.}
\tablenotetext{b}{Fractional uncertainty in the mean stacked flux, taken as the ratio of the 
error in the mean of the measured stacked flux ($\sigma/\sqrt{N}$) to the median measured flux, 
i.e., $\sigma (f^{meas})/(\sqrt{N}\langle f^{meas}\rangle)$.}
\label{tab:biaserr}
\end{deluxetable*}

\subsubsection{Exclusion Radius}

The most conservative measure of the median flux can be achieved by
stacking only those galaxies without any nearby sources.  The
full-width-half-max (FWHM) of the PSF at $100$\,$\mu$m is small enough
($\simeq 6\arcsec$) such that we can adopt an ``exclusion radius''
$r_{\rm exc}=$FWHM$/2$, and still have enough galaxies to stack.  We
first constructed a catalog containing all optical, $\ks$-band, and
IRAC detections in the GOODS-North field.  The $3$\,$\sigma$
sensitivities, as measured in a $2\arcsec$ diameter aperture in the
optical and $\ks$-band images, are ${\cal R}\simeq 27.6$ and
$\ks\simeq 24.55$.  The $3$\,$\sigma$ sensitivities of the GOODS-North
IRAC data, as measured in a $4\arcsec$ diameter aperture, are 26.56,
26.00, 24.17, and 24.16, for the 4 IRAC channels, respectively.  Any
galaxies in our sample that lie within a distance $r_{\rm exc}$ of any
source detected in the optical, near-IR, or with IRAC, are excluded
from the stack.  The sample itself was selected such that galaxies
that were confused with any nearby MIPS $24$\,$\mu$m sources were
excluded.

The median $100$\,$\mu$m fluxes obtained with and without adopting an
exclusion radius are listed in Table~\ref{tab:stacks}.  The
differences between these fluxes are smaller than the $1$\,$\sigma$
measurement uncertainties in the stacked fluxes.  The similarity in
flux between the exclusion and no-exclusion cases implies that any
objects that may cluster around the UV-selected galaxies do not
contribute significantly to the stacked far-infrared fluxes of the
UV-selected galaxies.  We note that there may exist very faint sources
that are undetected in the optical, near-IR, and IRAC data and which
may lie close to our targets.  However, we consider it unlikely for
sources that are faint at virtually all other wavelengths (optical,
near-IR, and the {\em Spitzer} IRAC and MIPS bands) to be bright
enough at $100$\,$\mu$m to contribute significantly to the stacked
flux of the UV-selected galaxies.  Performing a similar test at
$160$\,$\mu$m is not possible, as the larger FWHM results in an
exclusion radius that precludes any galaxies from being stacked.
Because the $100$ and $160$\,$\mu$m emission arises from the same
mechanism (dust emission), then one would conclude that the effects of
clustering are unlikely to be significant at $160$\,$\mu$m if the
$100$\,$\mu$m stacked emission yields similar infrared luminosities to
those obtained from the stacked $160$\,$\mu$m data.  In
Section~\ref{sec:luminosities} we present evidence that suggests that
an additional contribution to the $160$\,$\mu$m flux from other
sources must be negligible compared to the flux from the UV-selected
galaxies of interest here.  Based on this evidence, and to take
advantage of all the galaxies in our sample, we proceed by adopting
the fluxes obtained without using the exclusion radius; these values
are given in Table~\ref{tab:fluxes}.  Below, we discuss two additional
tests used to verify these stacked fluxes.

\subsubsection{Comparison with Random Stacks}

\begin{figure*}
\plotone{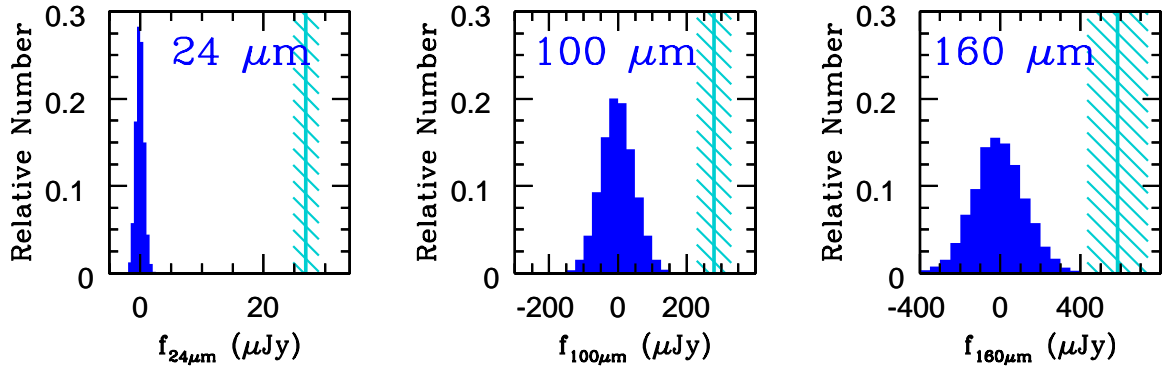}
\caption{Comparison of measured stacked fluxes (mean and uncertainty
  indicated by the solid lines and hashed regions, respectively) for
  Sample B and the distribution of fluxes obtained by stacking on
  $N=114$ random positions, repeated 10000 times (histograms).}
\label{fig:ranflux}
\end{figure*}

The probability of a chance measurement that results in fluxes as high
as the ones obtained by stacking on the positions of UV-selected
galaxies (i.e., the target stacks) can be determined by stacking at
random positions in the images (i.e., random stacks).  We performed
this random stack test on the radio and $24$, $100$, and $160$\,$\mu$m
data (we did not exclude positions that correspond to detected
sources).  The results for the mid-IR and IR simulations for Sample B
(stacking on $N=114$ positions) are shown in Figure~\ref{fig:ranflux}.
For this sample, the $24$ and $100$\,$\mu$m fluxes measured for 10,000
random stacks were never as high as those obtained for the target
stacks.  For the $160$\,$\mu$m data, 15 out of 10,000 random stacks
result in fluxes that were within the $1$\,$\sigma$ measurement
uncertainty of the target stack.  The random stacks indicate a very
low probability ($\la 0.2\%$) of accidentally recovering stacked
fluxes as high as the ones observed for the sample of $L^{\ast}$
galaxies at $z\sim 2$, and these probabilities are consistent with
those expected based on the measurement uncertainties of the stacked
fluxes.

\begin{figure*}
\plotone{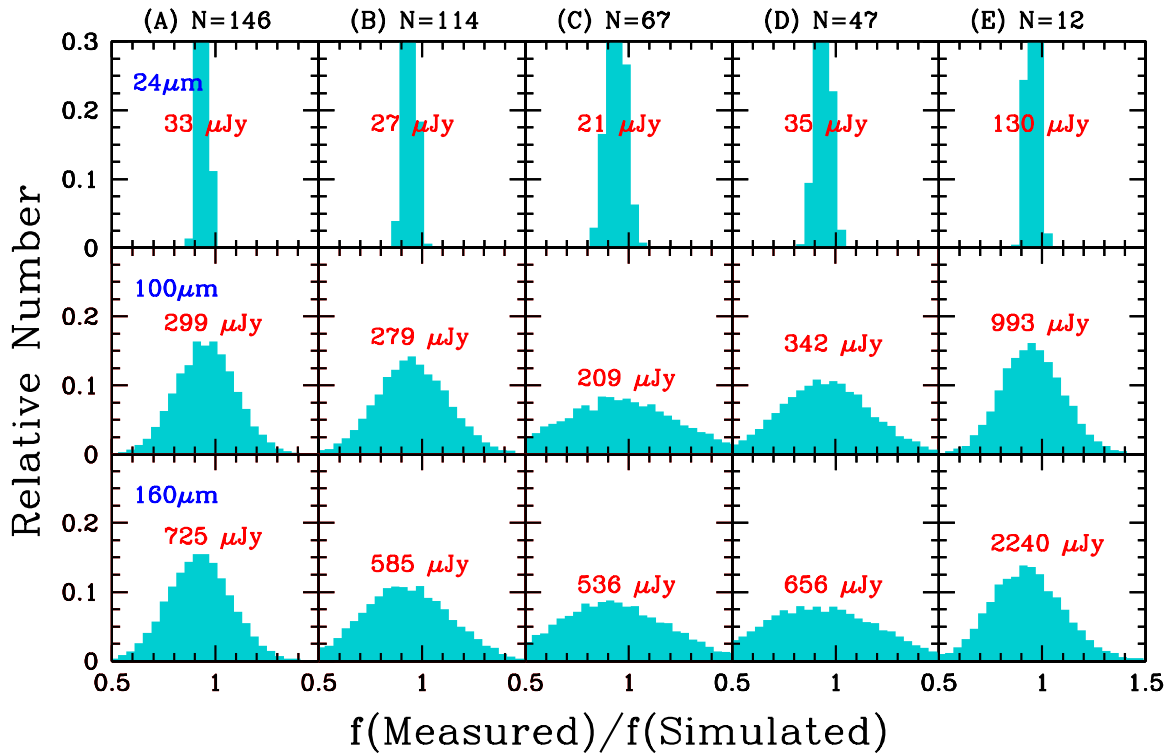}
\caption{Distribution of stacked fluxes for artificial galaxies added
  to the $24$, $100$, and $160$\,$\mu$m images (histograms) for Samples
  A through E, relative to the simulated flux (numbers in panels).}
\label{fig:simflux}
\end{figure*}

\subsubsection{Stacks of Simulated Galaxies}

Building on the random stack tests, point sources of known flux
density were added at random locations in the residual images (using
the same PSFs that are used to obtain photometry\footnote{We performed
  another test to determine if asymmetries in the intrinsic {\em
    Herschel} PSF result in systematic differences in the photometry
  obtained for simulated sources.  For this test, we redid the
  simulations by adding point sources of known flux density, where we
  assumed the flux profile given by the PACS PSF measured by observing
  the asteroid Vesta.  The sources were then recovered using the PSF
  measured from the GOODS-N {\em Herschel} images.  Based on this
  second set of simulations, we find less than $5\%$ systematic offset
  between the difference of input and output flux relative to the
  difference obtained when assuming the same PSF, both for adding
  sources to the images and recovering their fluxes.}) and were
recovered by a stacking analysis.  We added the same number of point
sources to the residual images as are used in constructing the stacks
for Samples A through E.  Flux densities were assigned based on a
Gaussian distribution with a mean equivalent to the median stacked
flux for each Sample (the stacks are insensitive to the dispersion in
simulated fluxes).  The ratio of the measured stack to simulated mean
flux are shown in Figure~\ref{fig:simflux} for the different samples,
with numbers and input mean fluxes indicated, for the $24$, $100$, and
$160$\,$\mu$m data.  This ratio is close to, but not exactly, equal to
$1$, with a bias of just a few percent.  The biases and dispersions
measured for the different stacks are summarized in
Table~\ref{tab:biaserr}.  The error in the mean recovered stacked flux
from the simulations is much smaller than the measurement uncertainty.
Therefore, for the subsequent discussion we assume a flux uncertainty
that is equal to the measurement uncertainty of the stacked flux.
Finally, we corrected the observed fluxes (listed in
Table~\ref{tab:fluxes}) by the bias factors given in
Table~\ref{tab:biaserr} before inferring the total infrared
luminosities, as we discuss in the next section.

\subsubsection{Summary of Tests}

We have performed several tests to validate our measures of the
stacked fluxes for UV-selected galaxies.  The method employed here is
able to recover the stacked fluxes of galaxies with a bias of just a
few percent relative to the known fluxes of artificial sources added
to the 1.4\,GHz, and $24$, $100$, and $160$\,$\mu$m images.  These
tests also imply a very low probability ($\la 0.2\%$) of accidentally
recovering stacked fluxes that are as high as the target stacked
fluxes.  Finally, we test for the effect of source clustering on the
stacked flux at $100$\,$\mu$m.  By excluding galaxies from the stack
which have any nearby sources, we find that the remaining galaxies
have a median flux that is identical within the measurement errors of
the median flux inferred for the entire sample of galaxies.  This
indicates that any sources that may cluster around the UV-selected
galaxies (if they exist) do not affect the stacking results.

\section{Infrared Luminosities and Dust SEDs}
\label{sec:luminosities}

\subsection{Dust SED Templates and Extrapolation to the Radio}

To infer the total infrared luminosities $L(8\,-\,1000\,\mu{\rm
  m})\equiv \lir$ of the $z\sim 2$ galaxies, we employed several
publicly available dust SED templates including those of
\citet{chary01} (``CE01''), \citet{dale02} (``DH02''), and
\citet{rieke09} (``Rieke+09'').  We also include the median templates
presented in \citet{elbaz11}.  Specifically, these authors define an
``infrared main sequence'' of galaxies, where the ratio of $\lir$ to
$\mirlum$ (IR8) is universal for most star-forming galaxies at $z\la
2.0$, having a value of $\lir/\mirlum = 4.9^{+2.9}_{-2.2}$.  The
``infrared starbursts'' are considered to be those galaxies with IR8
ratios in excess of $\approx 15$.  \citet{elbaz11} demonstrate that
variations in IR8 may be due primarily to differences in the infrared
luminosity surface density of galaxies, such that main sequence
galaxies are characterized by star formation that is more extended
than that present in starbursts (see Section~\ref{sec:discussion}).
We fit the median IR SED of these main sequence (``Elbaz+11-MS'') and
starburst galaxies (``Elbaz+11-SB'') to the stacked fluxes.

The \citet{reddy10a} calibration between rest-frame $8$\,$\mu$m
emission and infrared luminosity is considered as well.  In this case,
$\mirlum$ is computed by k-correcting the $24$\,$\mu$m flux using the
average of 12 local galaxy mid-IR SEDs specified in \citet{reddy06a}.
The correlation between $\mirlum$ and H$\alpha$ luminosity is then
used to infer the $\mirlum$--$\lir$ conversion, based on using
H$\alpha$ as an independent probe of the star formation
\citep{reddy10a}.  Finally, we make use of the radio-infrared
correlation to provide an independent estimate of $\lir$, assuming
that this relation does not evolve with redshift, as is consistent
with the evidence, at least at redshifts $z\la 3$ (e.g.,
\citealt{appleton04, ivison10, sargent10, bourne11, mao11}).  For the
subsequent analysis, we use the radio-infrared correlation published
in \citet{bell03}.  The sample of \citet{yun01} is roughly a factor of
10 larger than the one analyzed by \citet{bell03}.  However,
\citet{yun01} calibrate the $60$\,$\mu$m luminosity, $L_{\rm 60}$,
with radio luminosity, and the former requires some assumption about
the relation between $L_{\rm 60}$ and $\lir$.  To make the minimum
number of assumptions, we therefore adopted the \citet{bell03}
calibration which directly relates the total infrared luminosities to
the specific luminosity at $1.4$\,GHz.

The CE01 and Rieke+09 models are parameterized such that we can relate
any template to a given total infrared luminosity.  The Elbaz+11-MS/SB
templates are normalized to an infrared luminosity of
$\lir=10^{11}$\,L$_{\odot}$.  The DH02 models are presented as a
function of radiation field intensity and, following the literature
(e.g., \citealt{papovich07}), we associate infrared luminosities with
each of the DH02 templates assuming the empirical calibration of
\citet{marcillac06}.  The CE01 and Rieke+09 models also include an
extrapolation to radio wavelengths by assuming some version of the
local radio-far-infrared correlation \citep{condon92, yun01, bell03}.
For a consistent treatment, we have adjusted the models (or added to
them, in the case of DH02 and Elbaz+11-MS/SB) for a radio spectrum
with index $\gamma=-0.8$ \citep{condon92}, normalized to have a
specific luminosity at $1.4$\,GHz corresponding to the $\lir$ for that
template, assuming the \citet{bell03} calibration.

\subsection{Fitting Procedure}

Because the far-infrared peak of the dust emission in the SED
templates shifts to shorter wavelengths at higher luminosities, we can
either (1) fit these templates directly to the observed infrared and
radio fluxes (``luminosity-matched'' fitting), or (2) find the
template that best matches the infrared colors and then normalize this
template to the observed fluxes (``color-matched'' fitting).  The
color-matched fitting yields larger uncertainties in $\lir$ because
the color errors are larger than those for individual flux
measurements.  We have adopted both methods for comparison purposes.
The Elbaz+11-MS/SB templates are based on the composite IR spectral
energy distribution of the ``main sequence'' and ``starburst'' galaxy
samples of \citet{elbaz11}, and are normalized to
$\lir=10^{11}$\,L$_{\odot}$.  We assume the same spectral shape (i.e.,
no color dependence) when fitting these templates to the observed
fluxes.  Further, in finding the best-fit template, we include results
where all fluxes have been weighted equally, as well as being weighted
by their errors.  The equal weighting is done to ensure that the
higher S/N $24$\,$\mu$m data (and thus the smaller measurement
uncertainties at $24$\,$\mu$m) do not unduly skew the template fits.

\subsection{Infrared Luminosities}

Table~\ref{tab:lircomp} lists the $\lir$ corresponding to the best-fit
templates for different combinations of the stacked fluxes for Samples
A through F, after taking into account the biases and dispersions in these
fluxes (see discussion above and Table~\ref{tab:biaserr}).  Fitting
for different combinations of the infrared and radio data allows us to
determine if the inclusion of any of the stacked fluxes results in
significant changes in the inferred infrared luminosities.  Also
included in this table are the results from using the \citet{reddy10a}
calibration between $\mirlum$ and $\lir$, and the $\lir$ corresponding
radio luminosity assuming the \citet{bell03} radio-IR correlation.

\begin{deluxetable*}{llccccrc}
\tabletypesize{\scriptsize}
\tablewidth{0pc}
\tablecaption{Comparison of Infrared Luminosities ($\lir$)~\tablenotemark{a}}
\tablehead{
\colhead{Template} &
\colhead{$\lambda$, $\nu$~\tablenotemark{b}} &
\colhead{{\bf Sample A}} &
\colhead{{\bf Sample B}} &
\colhead{{\bf Sample C}} &
\colhead{{\bf Sample D}} &
\colhead{{\bf Sample E}} &
\colhead{{\bf Sample F}}}
\startdata
R10a~\tablenotemark{c} & 24                                     & $2.0\pm0.3$ & $1.8\pm0.1$ & $1.2\pm0.2$ & $2.7\pm0.2$ & $15.7\pm0.5$ & $0.3\pm0.1$\\
\\
Bell03~\tablenotemark{d} & 1.4                                  & $2.2\pm0.5$ & $2.2\pm0.4$ & $1.7\pm0.6$ & $2.9\pm0.8$ & $7.9\pm1.7$ & 3\,$\sigma$: $<3.0$ \\
\\
Elbaz+11-MS Lum~\tablenotemark{e} & 24                          & $1.9\pm0.2$ & $1.6\pm0.1$ & $1.2\pm0.1$ & $2.2\pm0.2$ & $9.0\pm0.3$ & $0.5\pm0.1$ \\
... & 100                                                       & $3.0\pm0.4$ & $2.9\pm0.5$ & $2.0\pm0.6$ & $3.8\pm0.7$ & $12.7\pm1.4$ & 3\,$\sigma$: $<2.8$ \\
... & 160                                                       & $3.0\pm0.4$ & $2.4\pm0.6$ & $2.1\pm0.6$ & $2.9\pm0.9$ & $11.1\pm1.7$ & $1.7\pm1.1$ \\
... & 100, 160                                                  & $3.0\pm0.3$ & $2.7\pm0.4$ & $2.1\pm0.4$ & $3.5\pm0.5$ & $12.0\pm1.0$ & 3\,$\sigma$: $<2.8$ \\
... & 24, 100, 160                                              & $2.3\pm0.3$ & $1.7\pm0.2$ & $1.3\pm0.2$ & $2.3\pm0.3$ & $9.2\pm0.4$ & 3\,$\sigma$: $<2.8$ \\
... & 24, 100, 160, 1.4                                         & $2.3\pm0.3$ & $1.7\pm0.2$ & $1.3\pm0.2$ & $2.3\pm0.3$ & $9.2\pm0.5$ & 3\,$\sigma$: $<2.8$ \\
Elbaz+11-MS Lum-Eq Weight~\tablenotemark{f} & 24, 100, 160, 1.4 & $3.0\pm0.2$ & $2.5\pm0.2$ & $2.1\pm0.2$ & $3.0\pm0.2$ & $11.3\pm0.3$ & 3\,$\sigma$: $<2.8$ \\
\\
Elbaz+11-SB Lum~\tablenotemark{e} & 24                          & $3.7\pm0.5$ & $3.1\pm0.2$ & $2.3\pm0.3$ & $4.3\pm0.4$ & $17.7\pm0.5$ & $1.0\pm0.2$ \\
... & 100                                                       & $2.4\pm0.3$ & $2.3\pm0.4$ & $1.6\pm0.5$ & $3.0\pm0.5$ & $10.0\pm1.1$ & 3\,$\sigma$: $<2.2$ \\
... & 160                                                       & $2.3\pm0.3$ & $1.9\pm0.5$ & $1.7\pm0.5$ & $2.3\pm0.7$ & $8.6\pm1.3$ & $1.3\pm0.9$ \\
... & 100, 160                                                  & $2.4\pm0.2$ & $2.1\pm0.3$ & $1.6\pm0.3$ & $2.7\pm0.4$ & $9.4\pm0.9$ & 3\,$\sigma$: $<2.2$ \\
... & 24, 100, 160                                              & $2.6\pm0.3$ & $2.7\pm0.3$ & $2.0\pm0.3$ & $3.6\pm0.4$ & $15.5\pm0.7$ & 3\,$\sigma$: $<2.2$ \\
... & 24, 100, 160, 1.4                                         & $2.6\pm0.4$ & $2.6\pm0.3$ & $2.0\pm0.4$ & $3.5\pm0.5$ & $15.0\pm0.8$ & 3\,$\sigma$: $<2.2$ \\
Elbaz+11-SB Lum-Eq Weight~\tablenotemark{f} & 24, 100, 160, 1.4 & $2.3\pm0.2$ & $2.0\pm0.1$ & $1.7\pm0.2$ & $2.4\pm0.2$ & $8.8\pm0.2$ & 3\,$\sigma$: $<2.2$ \\
\\
CE01 Lum~\tablenotemark{e} & 24                                 & $3.4\pm0.4$ & $2.6\pm0.2$ & $1.7\pm0.2$ & $4.1\pm0.3$ & $41.3\pm1.2$ & $0.4\pm0.1$ \\
... & 100                                                       & $2.4\pm0.3$ & $2.3\pm0.4$ & $1.6\pm0.5$ & $3.1\pm0.6$ & $10.9\pm1.2$ & 3\,$\sigma$: $<2.3$ \\
... & 160                                                       & $2.2\pm0.3$ & $1.8\pm0.4$ & $1.5\pm0.4$ & $2.1\pm0.6$ & $7.9\pm1.2$ & $1.2\pm0.8$ \\
... & 100, 160                                                  & $2.3\pm0.3$ & $2.0\pm0.3$ & $1.6\pm0.3$ & $2.7\pm0.4$ & $9.4\pm0.9$ & 3\,$\sigma$: $<2.3$ \\
... & 24, 100, 160                                              & $2.4\pm0.3$ & $2.3\pm0.3$ & $1.6\pm0.3$ & $3.3\pm0.4$ & $12.3\pm0.4$ & 3\,$\sigma$: $<2.3$ \\
... & 24, 100, 160, 1.4                                         & $2.4\pm0.4$ & $2.3\pm0.3$ & $1.6\pm0.4$ & $3.2\pm0.4$ & $12.1\pm0.6$ & 3\,$\sigma$: $<2.3$ \\
CE01 Lum-Eq Weight~\tablenotemark{f} & 24, 100, 160, 1.4        & $2.2\pm0.2$ & $1.8\pm0.1$ & $1.5\pm0.2$ & $2.3\pm0.2$ & $8.2\pm0.2$ & 3\,$\sigma$: $<2.3$ \\
CE01 Col~\tablenotemark{g} & 100, 160                           & $2.1\pm0.4$ & $1.8\pm0.5$ & $2.1\pm1.0$ & $1.7\pm0.6$ & $7.9\pm1.5$ & ... \\
... & 24, 100, 160                                              & $2.3\pm0.6$ & $2.2\pm0.8$ & $1.6\pm0.9$ & $3.0\pm1.2$ & $10.4\pm2.3$ & ... \\
CE01 Col-Eq Weight~\tablenotemark{h} & 24, 100, 160, 1.4        & $2.4\pm0.7$ & $2.2\pm0.8$ & $1.6\pm0.9$ & $3.0\pm1.2$ & $12.1\pm2.7$ & ... \\
\\
DH02 Lum~\tablenotemark{e} & 24                                 & $3.0\pm0.4$ & $2.3\pm0.2$ & $1.7\pm0.2$ & $3.6\pm0.3$ & $21.5\pm0.6$ & $0.6\pm0.1$ \\
... & 100                                                       & $2.1\pm0.3$ & $2.0\pm0.4$ & $1.5\pm0.5$ & $2.7\pm0.5$ & $7.8\pm0.9$ & 3\,$\sigma$: $<2.0$ \\
... & 160                                                       & $2.2\pm0.3$ & $1.8\pm0.4$ & $1.6\pm0.4$ & $2.1\pm0.6$ & $7.4\pm1.2$ & $1.3\pm0.9$ \\
... & 100, 160                                                  & $2.1\pm0.2$ & $1.9\pm0.2$ & $1.5\pm0.3$ & $2.5\pm0.4$ & $7.7\pm0.6$ & 3\,$\sigma$: $<2.0$ \\
... & 24, 100, 160                                              & $2.3\pm0.3$ & $2.2\pm0.2$ & $1.6\pm0.2$ & $3.0\pm0.4$ & $13.6\pm0.7$ & 3\,$\sigma$: $<2.0$ \\
... & 24, 100, 160, 1.4                                         & $2.3\pm0.3$ & $2.2\pm0.3$ & $1.6\pm0.3$ & $3.0\pm0.5$ & $13.2\pm0.8$ & 3\,$\sigma$: $<2.0$ \\ 
DH02 Lum-Eq Weight~\tablenotemark{f} & 24, 100, 160, 1.4        & $2.2\pm0.2$ & $1.8\pm0.1$ & $1.6\pm0.2$ & $2.2\pm0.2$ & $7.4\pm0.2$ & 3\,$\sigma$: $<2.0$ \\
DH02 Col~\tablenotemark{g} & 100, 160                           & $2.2\pm0.4$ & $1.5\pm0.5$ & $1.7\pm0.7$ & $1.7\pm0.6$ & $6.9\pm1.3$ & ... \\
... & 24, 100, 160                                              & $2.4\pm0.7$ & $2.1\pm0.7$ & $1.5\pm0.8$ & $2.8\pm1.1$ & $10.5\pm2.4$ & ... \\
DH02 Col-Eq Weight~\tablenotemark{h} & 24, 100, 160, 1.4        & $2.3\pm0.6$ & $2.2\pm0.8$ & $1.6\pm0.9$ & $3.6\pm1.5$ & $9.5\pm2.2$ & ... \\
\\
Rieke+09 Lum~\tablenotemark{e} & 24                             & $5.0\pm0.6$ & $3.6\pm0.3$ & $2.0\pm0.2$ & $6.1\pm0.5$ & $46.1\pm1.3$ & $0.5\pm0.1$ \\
... & 100                                                       & $2.5\pm0.3$ & $2.3\pm0.4$ & $1.8\pm0.6$ & $3.0\pm0.5$ & $8.2\pm0.9$ & 3\,$\sigma$: $<2.3$ \\
... & 160                                                       & $2.2\pm0.3$ & $1.9\pm0.4$ & $1.7\pm0.5$ & $2.1\pm0.6$ & $6.1\pm0.9$ & $1.4\pm1.0$ \\
... & 100, 160                                                  & $2.3\pm0.2$ & $2.1\pm0.2$ & $1.7\pm0.2$ & $2.5\pm0.3$ & $7.1\pm0.6$ & 3\,$\sigma$: $<2.3$ \\
... & 24, 100, 160                                              & $2.3\pm0.2$ & $2.3\pm0.3$ & $1.8\pm0.3$ & $3.1\pm0.4$ & $11.2\pm0.7$ & 3\,$\sigma$: $<2.3$ \\
... & 24, 100, 160, 1.4                                         & $2.3\pm0.3$ & $2.3\pm0.3$ & $1.8\pm0.4$ & $3.0\pm0.5$ & $11.0\pm0.9$ & 3\,$\sigma$: $<2.3$ \\
Rieke+09 Lum-Eq Weight~\tablenotemark{f} & 24, 100, 160, 1.4    & $2.2\pm0.2$ & $1.9\pm0.1$ & $1.7\pm0.2$ & $2.2\pm0.2$ & $6.3\pm0.2$ & 3\,$\sigma$: $<2.3$ \\ 
Rieke+09 Col~\tablenotemark{g} & 100, 160                       & $3.1\pm0.6$ & $2.8\pm0.8$ & $2.4\pm1.3$ & $3.6\pm1.3$ & $12.5\pm2.7$ & ... \\
... & 24, 100, 160                                              & $2.7\pm0.7$ & $2.4\pm0.8$ & $1.8\pm1.0$ & $3.2\pm1.3$ & $11.9\pm2.7$ & ... \\
Rieke+09 Col-Eq Weight~\tablenotemark{h} & 24, 100, 160, 1.4    & $2.6\pm0.7$ & $2.3\pm0.8$ & $1.8\pm0.9$ & $3.1\pm1.2$ & $12.1\pm2.7$ & ... \\
\enddata
\tablenotetext{a}{\,Luminosities are in units of $10^{11}$\,L$_{\odot}$.  Errors in luminosities are derived from
the uncertainties in the stacked fluxes (see text).}
\tablenotetext{b}{Wavelengths (in micron) or frequency (in GHz) used to compute the best-fit $\lir$.}
\tablenotetext{c}{$\lir$ determined from the calibration of $24$\,$\mu$m with $\lir$ of \citet{reddy10a}, which is
based upon the correlation between $8$\,$\mu$m and H$\alpha$ luminosity for $L^{\ast}$ galaxies
at $z\sim 2$.}
\tablenotetext{d}{$\lir$ determined from the radio-IR correlation of \citet{bell03}, and assuming a radio
spectral slope of $\gamma=-0.8$.}
\tablenotetext{e}{$\lir$ for the \citet{elbaz11} main sequence/starburst, \citet{chary01}, \citet{dale02}, or \citet{rieke09} template that best matches 
the observed fluxes (weighted by the flux errors), irrespective of the infrared colors.}
\tablenotetext{f}{$\lir$ for the \citet{elbaz11} main sequence/starburst, \citet{chary01}, \citet{dale02}, or \citet{rieke09} template that best matches 
the observed fluxes (including 1.4\,GHz), with all fluxes given equal weight in the fitting, irrespective of the 
infrared colors.}
\tablenotetext{g}{$\lir$ for the \citet{chary01}, \citet{dale02}, or \citet{rieke09} template that best fits 
the observed infrared colors and is normalized to match the observed fluxes, with colors and fluxes weighted by
their errors.}
\tablenotetext{h}{$\lir$ for the \citet{chary01}, \citet{dale02}, or \citet{rieke09} template that best fits
the observed infrared colors and is normalized to match the observed fluxes (including 1.4\,GHz), with colors
and fluxes weighted equally.}
\label{tab:lircomp}
\end{deluxetable*}

The systematic uncertainties in $\log(\lir)$ for the different
combinations of templates, fluxes, and fitting methods for each of the
subsamples are typically within $0.1$\,dex.  We note that the $\lir$
derived from $160$\,$\mu$m is not systematically larger than that inferred from
$100$\,$\mu$m, and hence runs counter to what one might expect if the
larger $160$\,$\mu$m beam included more sources clustered around the
UV-selected galaxies that contribute to the measured far-infrared
flux.  The stacked radio data have significantly higher resolution
($\sim 1\farcs7$ at $1.4$\,GHz versus $\sim 11\arcsec$ at
$160$\,$\mu$m) and yield a radio flux that implies an $\lir$ that is
also not any smaller than the value inferred from the $160$\,$\mu$m
data.  Based on this evidence, we conclude that the $160$\,$\mu$m
emission from sources proximate to the UV-selected galaxies is
negligible compared to the emission from the UV-selected galaxies
themselves.

\begin{figure}
\plotone{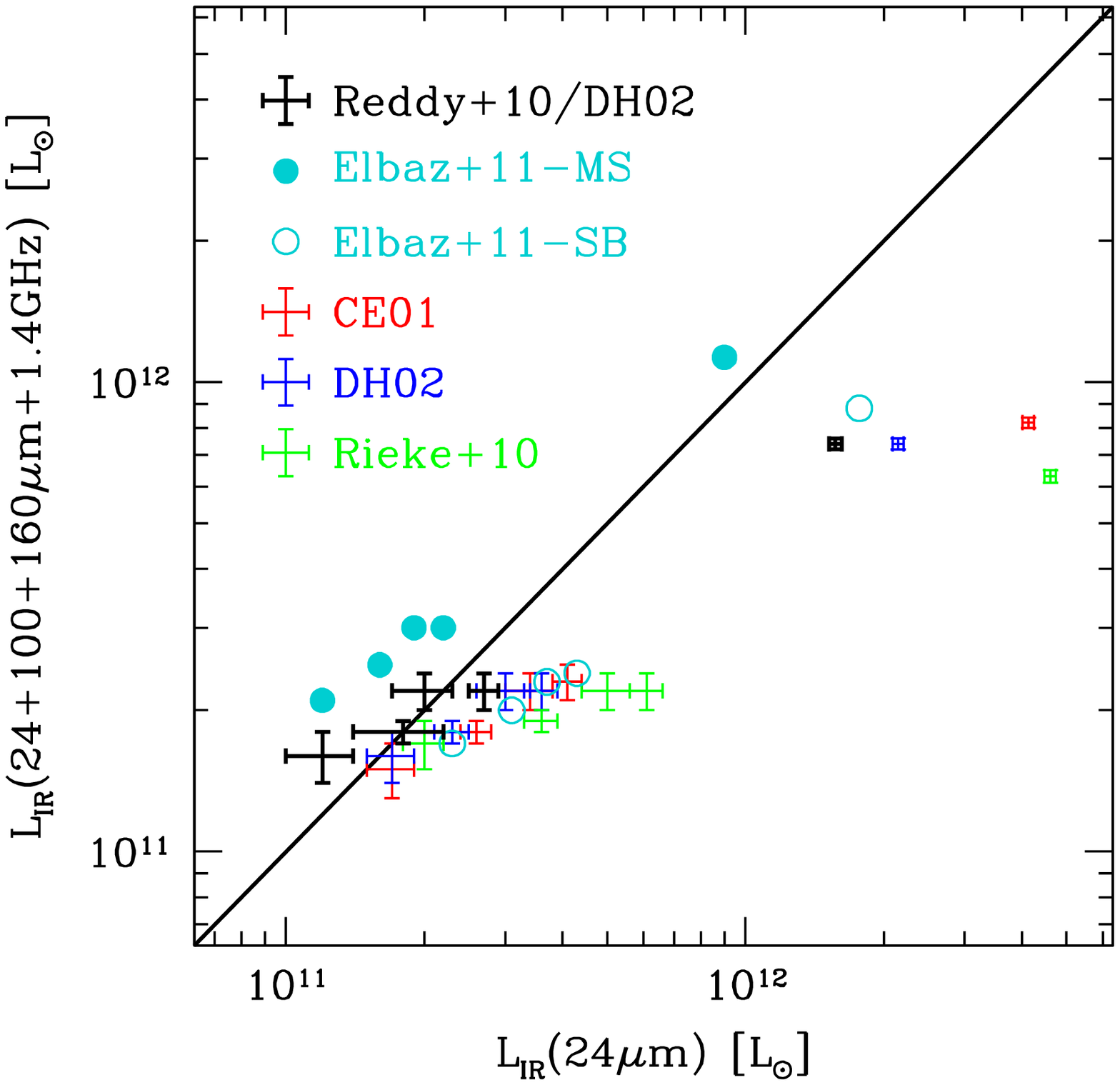}
\caption{Comparison between the $\lir$ computed from the $24$, $100$,
  $160$\,$\mu$m, and $1.4$\,GHz equally-weighted fluxes, and the
  $\lir$ computed from the $24$\,$\mu$m data only.  The heavy black
  points denote the $\lir$ computed from the \citet{reddy10a}
  calibration versus the $\lir$ obtained from the best-fit DH02
  template that includes all the data ($24$, $100$, $160$\,$\mu$m, and
  1.4GHz), equally-weighted.  Error bars reflect the uncertainty in
  the $\lir$ inferred from the stacked measurements, except for the
  high luminosity subsample (Sample E), where the errors include the
  intrinsic dispersion in $24$\,$\mu$m fluxes of objects that
  contribute to the sample.  For clarity, error bars are not shown for
  the Elbaz+11-MS/SB points.}
\label{fig:lirvlir}
\end{figure}

The notable outliers listed in Table~\ref{tab:lircomp} are the $\lir$
determined from $24$\,$\mu$m flux density alone, which tend to
overpredict $\lir$, particularly for the highest luminosity subsamples
D and E, relative to cases where we include the $100$ and
$160$\,$\mu$m data in most of the template fits
(Figure~\ref{fig:lirvlir}).  Alternatively, with the Elbaz+11-MS
template, $24$\,$\mu$m-only determinations result in systematically
lower $\lir$.  These differences are due to the intrinsic variation in
the ratio of $\mirlum$ to $\lir$ luminosity present in the templates
relative to the observed SED.  The systematic overestimation of $\lir$
based on $\mirlum$ alone has been noted before for ultraluminous
infrared galaxies studied with {\em Spitzer} \citep{papovich07,
  daddi07a, papovich09, murphy11, magnelli11} and {\em Herschel}
\citep{nordon10, elbaz10}.  For the samples considered here, the same
$24$\,$\mu$m bias is apparent for all templates (except for
Elbaz+11-MS) even at LIRG luminosities, where the difference between
the best-fit $\lir$ and that determined from $24$\,$\mu$m alone is
similar to the $1$\,$\sigma$ uncertainties in $\lir$.

Using {\em Spitzer} and {\em Herschel} data for galaxies at redshifts
$z\la 2.0$, \citet{elbaz11} investigate the physical reasons for the
overestimation of $\lir$ based on rest-frame $\mirlum$ alone.  These
authors point out that the SED templates used to infer $\lir$ are
calibrated to match local ULIRGs, which are starbursts with compact,
high surface density star formation.  This contrasts with ULIRGs at
high redshift ($z\ga 2.0$) which have more extended star formation
occuring over longer timescales.  These differences in star formation
surface densities and timescales lead to noticeable differences in the
IR SEDs because of the variations in the spatial distribution of dust
with respect to the massive stars that are heating this dust.  In
Section~\ref{sec:discussion}, we discuss the $\lir/\mirlum$ ratios found
here and place them in the context of the ratios found for other
star-forming galaxies with similar luminosities but at lower
redshifts.

We conclude by noting that the \citet{reddy10a} calibration, which is
based on the correlation between $\mirlum$ and dust-corrected
H$\alpha$ luminosity at $z\sim 2$, predicts $\lir$ for LIRGs that are
similar within the errors to those computed using the combined {\em
  Spitzer}, {\em Herschel}, and VLA data.  Finally, the infrared
luminosities based on the radio data alone are in excellent agreement
with the infrared luminosities inferred from fitting the dust
templates to the observed $24$, $100$, and $160$\,$\mu$m fluxes.

\subsection{Comparison of Dust SEDs}

\begin{figure}
\plotone{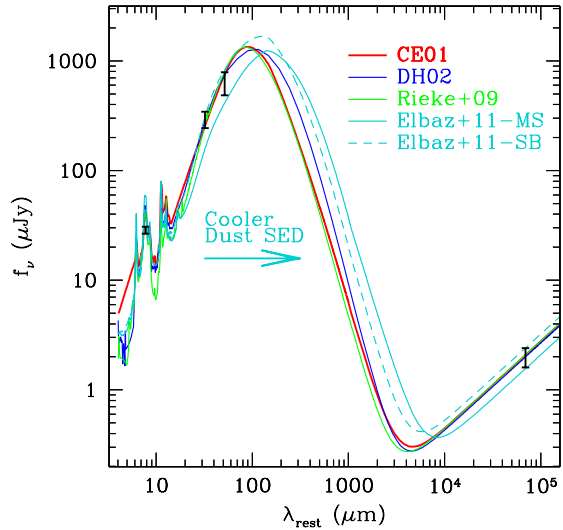}
\caption{Comparison of the \citet{chary01}, \citet{dale02},
  \citet{rieke09}, and \citet{elbaz11} template fits to the observed
  $24$, $100$, and $160$\,$\mu$m and $1.4$\,GHz measurements, derived
  by scaling the template that best matches the observed fluxes, for
  $L^{\ast}$ galaxies at $z\sim 2$.  The total infrared luminosities
  for the three templates are identical within the uncertainties
  (Table~\ref{tab:lircomp}).}
\label{fig:sedcomp}
\end{figure}

Figure~\ref{fig:sedcomp} compares the dust SEDs found for the
luminosity-matched fitting to the $24$, $100$, and $160$\,$\mu$m, and
1.4\,GHz data, weighted by their errors.  The dust SEDs for the
different models are broadly consistent with each other; one
difference is in the stronger silicate absorption at $9$\,$\mu$m and
additional aromatic features longward of $10$\,$\mu$m in the Rieke+09
model fit.  The Rieke+09 templates are calibrated using {\em
  Spitzer}/IRS spectra, rather than photometry, between $5$ and
$36$\,$\mu$m, and will understandably include mid-IR spectral features
that are not present in the other models.

The most notable difference between the templates can be seen in
Figure~\ref{fig:sedcomp}: the Elbaz+11 and DH02 templates exhibit a
broader range of dust temperatures with a colder dust component,
relative to CE01 and Rieke+09.  Resolving the full shape of the SED,
and the average dust temperatures, will require larger stacked samples
and deeper data in the submillimeter and millimeter regime.  In any
case, while some obvious differences remain between the model
templates given the data at our disposal, the integrals of these SEDs,
namely the total $\lir$, are essentially identical within the
uncertainties (Table~\ref{tab:lircomp}).  For the subsequent analysis,
we assume the $\lir$ determined from the DH02 model that best fits the
error-weighted fluxes at 1.4\,GHz, $24$, $100$, and $160$\,$\mu$m.
Assuming the $\lir$ derived using any of the other models does not
affect our conclusions.

\begin{figure*}
\plotone{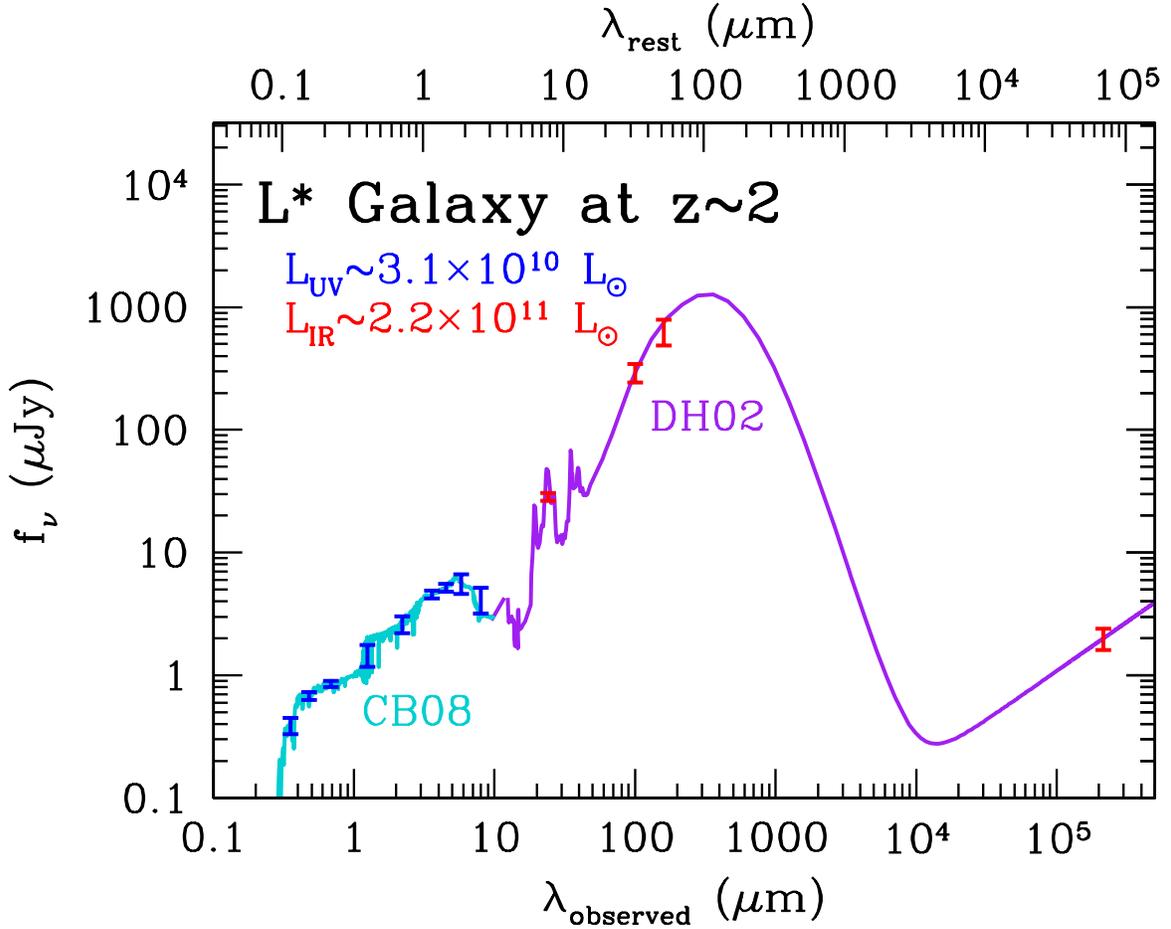}
\caption{Best-fit stellar population (CB08) and dust (DH02) SEDs for
  typical $L^{\ast}$ galaxies at $z\sim 2$.  Included are $U_{\rm
    n}G{\cal R}$+$J\ks$, {\em Spitzer}/IRAC $3.6\,-\,8.0$\,$\mu$m,
  {\em Spitzer} MIPS $24$\,$\mu$m, {\em Herschel} PACS $100$ and
  $160$\,$\mu$m, and VLA 1.4\,GHz stacked measurements.}
\label{fig:avesed}
\end{figure*}

\subsection{The Dust SED of Typical Star-Forming Galaxies at $z\sim 2$}

The average stellar population and dust SEDs for $L^{\ast}_{\rm UV}$
galaxies at $z\sim 2$ are shown in Figure~\ref{fig:avesed}.  The
optical photometry indicates that an $L^{\ast}_{\rm UV}$ galaxy at
$z\sim 2$ has a UV luminosity of $\luv \simeq
3.1\times10^{10}$\,L$_{\odot}$ and the dust SED indicates a total
infrared luminosity of $\lir \simeq 2.2\times 10^{11}$\,L$_{\odot}$.
The {\em Herschel} and VLA data confirm the previous finding that
UV-selected galaxies at $z\sim 2$ with ${\cal R}<25.5$ are luminous
infrared galaxies (LIRGs; \citealt{reddy05a, reddy06a, reddy10a,
  adel00}); in particular, the median $\lir$ found here is virtually
identical to that found by \citet{reddy10a} based on an analysis of
the $24$\,$\mu$m, H$\alpha$, and UV emission for UV-selected galaxies
at $z\sim 2$.  We also point out that the median value of $\lir$ for
$L^{\ast}_{\rm UV}$ galaxies is similar (to within a factor of
$\approx 3$) to the value of $L^{\ast}_{\rm IR}$ deduced from direct
{\em Spitzer} determinations of the IR LF at $z\sim 2$
\citep{magnelli11}.

\begin{figure}
\plotone{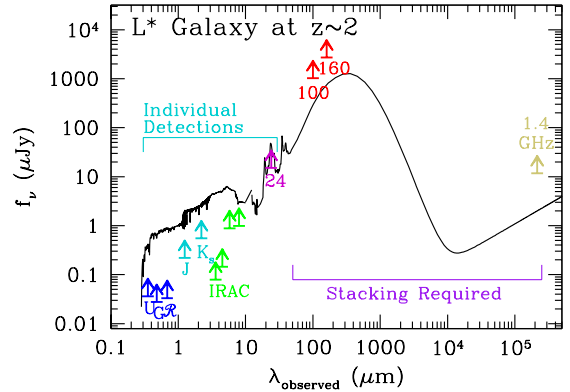}
\caption{Detections limits ($3$\,$\sigma$) for the ground-based optical
  ($U_{\rm n}G{\cal R}$), ground-based near-IR ($J\ks$), {\em
    Spitzer}/IRAC ($3.6\,-\,8.0$\,$\mu$m), {\em Spitzer}/MIPS
  ($24$\,$\mu$m), {\em Herschel}/PACS (100, 160\,$\mu$m), and VLA
  1.4\,GHz data in the GOODS-North field, relative to the SED of an
  $L^{\ast}$ galaxy at $z\sim 2$.}
\label{fig:detlims}
\end{figure}

The necessity of stacking the UV-selected galaxies, even for
GOODS-depth {\em Herschel} PACS data, is illustrated in
Figure~\ref{fig:detlims}.  Typical $L^{\ast}$ galaxies at $z\sim 2$
are easily detected at wavelengths blueward of observed $24$\,$\mu$m
given the depths of the data considered here.  Redward of this point,
however, the dust emission is not sufficient for directly detecting
these galaxies, thus stacking in the infrared and radio bands is
required.  Fortunately, the {\em Herschel} PACS sensivity and
resolution are such that we can for the first time significantly
detect the average thermal emission of non-lensed $L^{\ast}$ galaxies
at $z\sim 2$.

\subsection{Variation of $\lir$ with UV Slope and Bolometric Luminosity}

The stacks for the different subsamples indicate that galaxies with
bluer UV spectral slopes are on average less infrared luminous than
those with red UV spectral slopes (Table~\ref{tab:lircomp}).  This
systematic effect is approximately equal in magnitude to the
uncertainties in the stacked flux measurements.  Not surprisingly,
those galaxies that are inferred to have $\lbol \ge
10^{12}$\,L$_{\odot}$ based on the MIPS $24$\,$\mu$m data (Sample E;
Table~\ref{tab:lircomp}) have median stacked fluxes at $100$ and
$160$\,$\mu$m that are a factor of $3-4$ times larger than the
corresponding fluxes for the $L^{\ast}$ sample.  The median infrared
luminosity for Sample E is not as large as that computed from
$24$\,$\mu$m alone and hence not as large as the value of $\lbol$ used
to construct this subsample (e.g., Figure~\ref{fig:lirvlir}).
However, the best-fit template to the mid-IR, IR, and radio fluxes
indicates that galaxies in Sample E are still more infrared luminous
than galaxies in other subsamples.  These galaxies correspond to
low-luminosity ULIRGs or higher luminosity LIRGs.  In the next
section, we compare the infrared and UV luminosities for each of the
subsamples of UV-selected galaxies.

\section{Discussion}
\label{sec:discussion}

In the following, we discuss our results on $L^{\ast}$ galaxies at
$z\sim 2$ in the context of the infrared properties of star-forming
galaxies with similar luminosities at lower redshifts.  We then
discuss the combined UV and IR luminosity measurements for $z\sim 2$
galaxies and the implication for their average dust attenuation.  We
also compare the measured dust attenuation with that inferred from the
local correlation between dustiness and UV slope.  Finally, we use the
{\em Herschel} data to determine the median bolometric luminosities
and star formation rates of galaxies in our sample, the correlation
between bolometric luminosity and dust attenuation, and the variation
in dust attenuation with UV luminosity.

\subsection{Ratio of Infrared to Mid-Infrared Luminosity}

For a consistent comparison with \citet{elbaz11}, we have recomputed
$\mirlum$ using the mid-IR SED of M82 to {\em k}-correct the
$24$\,$\mu$m flux.  The implied {\em k}-corrections are not
substantially different than those obtained from the average mid-IR
SED of the 12 local star-forming galaxies listed in \citet{reddy06a}.
Adopting the CE01 luminosity matched and weighted value of $\lir$, we
compute IR8\,$= 7.7\pm1.6$, $8.9\pm1.3$, $8.3\pm2.4$, $8.9\pm1.4$, and
$8.4\pm0.5$, for Samples A, B, C, D, and E, respectively.  Assuming
the $\lir$ computed from the $100$ and $160$\,$\mu$m data only results
in ratios that are not significantly different than the ones quoted
above.  These values imply that the galaxies in our sample
predominantly lie on the boundary between the ratios found for
infrared main sequence galaxies (IR8$=4.9^{+2.9}_{-2.2}$), and those
found for starburst galaxies which have an exponential tail
distribution extending to IR8\,$=15\,-\,20$.  The relatively high IR8
ratios also may explain why the typically-adopted templates (e.g.,
CE01) tend to not fail as badly for galaxies in our sample relative to
more luminous ULIRGs at $z\sim 2$ when inferring $\lir$ from $\mirlum$
alone (e.g., \citealt{reddy06a, reddy10a}).

A detailed comparison between the morphologies of the $z\sim 2$
galaxies studied here and local star-forming galaxies is beyond the
scope of this paper.  However, we can still make some general
inferences regarding the degree of ``compactness'' of the IR emission
in these galaxies and how it may affect their IR8 ratios.  We compute
the IR luminosity surface density following \citet{elbaz11}:

\begin{eqnarray}
\Sigma_{\rm IR} \equiv \frac{\lir/2}{\pi r^2_{\rm IR}}.
\end{eqnarray}

The typical UV half-light radius of galaxies in our sample is $r_{\rm
  UV}\simeq 2$\,kpc.  If we assume that the IR half-light radius is
roughly half this value, as suggested by \citet{elbaz11}, then $r_{\rm
  IR}\approx 1$\,kpc, implying $\Sigma_{\rm IR} \approx 3\times
10^{10}$\,L$_\odot$\,kpc$^{-2}$.  On the other hand, if $r_{\rm IR}
\approx r_{\rm UV}$ as might be expected for galaxies at high redshift
where the UV emission is dominated by OB stars, then $\Sigma_{\rm IR}
\approx 8\times 10^{9}$\,L$_\odot$\,kpc$^{-2}$.  Even this lower value
is a factor of 4 larger than the $\Sigma_{\rm IR} \la 2\times
10^{9}$\,L$_\odot$\,kpc$^{-2}$ typical of the infrared main sequence
galaxies with extended star formation in the \citet{elbaz11} sample.
The relevance for the present discussion is that while $L^{\ast}$
galaxies at $z\sim 2.3$ appear to have IR8 ratios that are a factor of
$\approx 2$ larger than for main sequence galaxies at lower redshifts,
they also exhibit $\Sigma_{\rm IR}$ that are larger than those found
for main sequence galaxies, and hence are more compact for their IR
luminosity.  Formally, galaxies in our sample would lie on the
``boundary'' between IR main sequence and IR starburst galaxies.
These larger IR8 ratios result in the over-prediction of $\lir$ based
on $\mirlum$ alone when using the standard templates, as noted in
Section~\ref{sec:luminosities}.

One possibility is that the IR8 ratio for typical star-forming
($L^{\ast}$) galaxies may evolve with redshift, transitioning from
IR8\,$\simeq 4.9$ at $z\la 2.0$ to IR8\,$\simeq 8$ at $z\sim 2.3$.
This effect is likely due to the smaller sizes of high redshift
galaxies for their IR luminosities relative to lower redshift
galaxies.  Finally, it is worth noting that the {\em star formation
  rate} surface density appears to be roughly constant for $L^{\ast}$
galaxies at higher redshifts ($z\sim 4\,-\,7$; $\Sigma_{\rm SFR}
\simeq 1.9$\,M$_\odot$\,yr$^{-1}$\,kpc$^{-2}$; e.g.,
\citealt{oesch10}), suggesting that there may be a similar commonality
in the IR8 ratios (if one could measure them) for typical star-forming
galaxies at very high redshift, just as is observed for main sequence
galaxies at $z\la 2.0$ \citep{elbaz11}.

\subsection{Dust Obscuration of Typical Star-Forming Galaxies at $z\sim 2$}

\subsubsection{Definitions Relevant to Dust Obscuration}

Before proceeding, it is useful to define several terms that have been
typically used interchangeably in the literature.  First, we define
``dust obscuration'', or attenuation, as $\lir/\luv$.  Note that this
ratio is not equivalent to the ratio of obscured to unobscured star
formation rate, SFR$_{\rm IR}$/SFR$_{\rm UV}$, given the difference in
scaling required to convert the UV and IR luminosities to star
formation rates.  We also define the ``dust correction factor'' needed
to recover the total star formation rate from that computed based on
the unobscured UV luminosity as (SFR$_{\rm IR}$+SFR$_{\rm
  UV}$)/SFR$_{\rm UV}$ $\equiv 1 +$ SFR$_{\rm IR}$/SFR$_{\rm UV}$.
For the $L^{\ast}$ sample, the median dust obscuration is $\lir/\luv =
7.1\pm1.1$, the ratio of obscured to unobscured SFR is SFR$_{\rm
  IR}$/SFR$_{\rm UV} = 4.2\pm0.6$, and the dust correction factor is
$5.2\pm 0.6$.  The $\luv$, $\lir$, ratio of obscured to unobscured
SFR, and total SFR for each subsample are listed in
Table~\ref{tab:lums}.  For the conversion to SFR, we assume a
\citet{salpeter55} initial mass function with limits from 0.1 to
100\,M$_\odot$ and the \citet{kennicutt98} conversions betweeen UV/IR
luminosity and SFR.  The results indicate that roughly $80\%$ of the
star formation is obscured for $L^{\ast}$ galaxies at $z\sim 2$.

\begin{deluxetable*}{lcrcc}
\tabletypesize{\scriptsize}
\tablewidth{0pc}
\tablecaption{Properties of the Stacks IV: Derived Quantities}
\tablehead{
\colhead{} &
\colhead{$\luv$~\tablenotemark{a}} &
\colhead{$\lir$~\tablenotemark{b}} &
\colhead{} &
\colhead{SFR(UV)+SFR(IR)~\tablenotemark{d}} \\
\colhead{Sample} &
\colhead{($10^{11}$\,L$_\odot$)} &
\colhead{($10^{11}$\,L$_\odot$)} &
\colhead{1+SFR(IR)/SFR(UV)~\tablenotemark{c}\tablenotemark{d}} &
\colhead{(M$_\odot$\,yr$^{-1}$)}}
\startdata
{\bf A.} & $0.32\pm0.02$ & $2.3\pm0.3$ & $5.3\pm0.6$ & $49\pm6$ \\
{\bf B.} & $0.31\pm0.02$ & $2.2\pm0.3$ & $5.2\pm0.6$ & $47\pm6$ \\
{\bf C.} & $0.33\pm0.03$ & $1.6\pm0.3$ & $3.9\pm0.6$ & $37\pm6$ \\
{\bf D.} & $0.28\pm0.02$ & $3.0\pm0.5$ & $7.2\pm1.1$ & $60\pm9$ \\
{\bf E.} & $0.35\pm0.07$ & $13.2\pm0.8$ & $20.9\pm3.2$ & $238\pm 15.5$ \\
{\bf F.} & $0.38\pm0.05$ & 3\,$\sigma$: $<2.0$ & 3\,$\sigma$: $<2.4$ & 3\,$\sigma$: $<58$ \\
\enddata
\tablenotetext{a}{Mean and error in mean of UV luminosity in 
units of $10^{11}$\,L$_\odot$.}
\tablenotetext{b}{Infrared luminosity, in units of $10^{11}$\,L$_\odot$, 
derived from color-matching and normalizing the \citet{dale02} models 
to the observed fluxes.  For Sample F, we assume the upper limit in $\lir$
implied by the observed fluxes at $24$ and $160$\,$\mu$m and the upper limit
at $100$\,$\mu$m and $1.4$\,GHz.}
\tablenotetext{c}{Dust correction factor required to recover the total SFR
from the UV-determined SFR.}
\tablenotetext{d}{Median star formation rates in M$_\odot$\,yr$^{-1}$
assuming a \citet{salpeter55} IMF from $0.1$ to $100$\,M$_\odot$ and the 
\citet{kennicutt98} relations between UV/IR luminosity and star 
formation rate.  For the ``young'' subsample (Sample F), we multiply the
UV SFR determined from the \citet{kennicutt98} relation by a factor of 2.
This is done to account for the fact that the mix
of O and B stars contributing to the UV continuum emission has not equilibrated
for ages $\la 100$\,Myr (assuming a constant star formation); thus, the \citet{kennicutt98}
conversion between UV luminosity and SFR will underpredict the total SFR
for such ``young'' galaxies.}
\label{tab:lums}
\end{deluxetable*}

The dust obscuration varies between $\lir/\luv = 4.8\pm1.0$ for the
subsample with blue UV slopes (Sample C) and $\lir/\luv=10.7\pm1.9$
for the subsample with red UV slopes (Sample D), and is as high as
$\lir/\luv \approx 37.7\pm7.9$ for the most bolometrically-luminous
galaxies (Sample E).  These observations imply a trend in dustiness
with both UV slope and bolometric luminosity.

\subsubsection{Validity of the UV Attenuation Curve for $L^{\ast}$ Galaxies
at $z\sim 2$}

\begin{figure}
\plotone{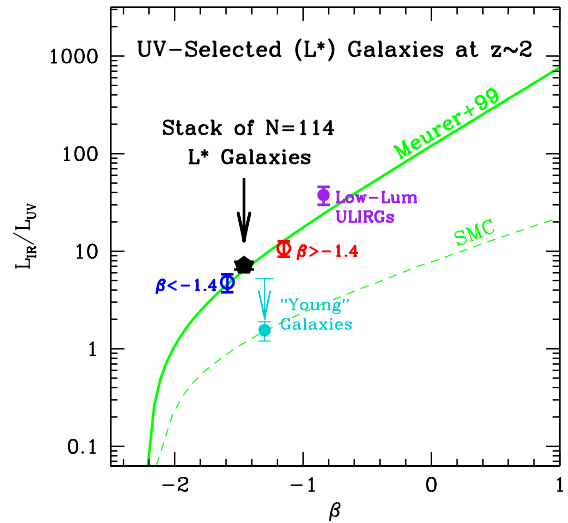}
\caption{Mean dust attenuation ($\lir/\luv$) versus UV slope ($\beta$)
  for different subsamples of $z\sim 2$ galaxies.  Also shown are
  attenuation curves for the SMC and for local UV starbursts from
  \citet{meurer99}, and the $3$\,$\sigma$ upper limit and stacked
  $24$\,$\mu$m implied value ({\em cyan} point) of the dust
  attenuation for the youngest galaxies in our sample.}
\label{fig:ebmv100}
\end{figure}

In particular, we show the dust obscuration derived for these samples
as a function of UV slope, $\beta$, in Figure~\ref{fig:ebmv100}.  Up
to luminosities of $\lir \approx 10^{12}$\,L$_{\odot}$, we find that
galaxies with redder $\beta$ are on average dustier.  Furthermore, the
correlation between dustiness and UV slope is essentially identical to
that found for local starburst galaxies \citep{meurer99, calzetti00}.
This result has been found by several other investigations targeting
moderately luminous galaxies and using a variety of star formation
tracers at $z\sim 2$ (e.g., \citealt{reddy04, reddy06a, reddy10a,
  daddi07a, pannella09}) and $z\sim 3$ (e.g., \citealt{seibert02,
  nandra02, magdis10a, magdis10b}).

Infrared selection, e.g., such as $24$\,$\mu$m selection, generally
results in samples where the bulk of galaxies do not abide by the
\citet{meurer99} or \citet{calzetti00} attenuation curves (e.g.,
\citealt{murphy11}).  As discussed in \citet{reddy06a} and
\citet{reddy10a}, the correlation between UV slope and dust
attenuation breaks down for more infrared luminous galaxies at $z\sim
2$, as well as younger galaxies with ages $\la 100$~Myr at the same
redshifts, where the latter tend to follow a steeper attenuation curve
(i.e., they are less reddened at a given UV slope than predicted by
the \citealt{meurer99} relation).  Systematic deviations from the
local starburst attenuation curve have also been observed at lower
redshift ($z\la 2$) based on {\em Herschel}/PACS and SPIRE data (e.g.,
\citealt{buat10, burgarella11}) and {\em Akari} data \citep{buat11}.

Much of the aforementioned deviation from the local starburst
attenuation curve can be understood in the context of the range of
bolometric luminosity probed by the different UV and IR selections.
UV color selection is sensitive to galaxies with moderate ($L^{\ast}$)
luminosities and lower dust extinction than those selected in the
infrared.  Because dust attenuation is a strong function of bolometric
luminosity and the validity of the \citet{meurer99} relation is
luminosity dependent \citep{meurer99, goldader02, reddy06a, reddy10a},
it is natural to expect departures from this relation for galaxies
that may be selected via their infrared emission.  Deviations may also
be observed in infrared luminous galaxies that also have large stellar
masses; in this case, the UV continuum associated with the massive OB
stars may be extinguished relative the UV emission from less massive
stars, resulting in a redder UV continuum slope for a given dust
obscuration (e.g., \citealt{murphy11, buat10}).  These results stress
that one must take care in applying {\em any} starburst attenuation
curve to high redshift galaxies, with the realization that such
relations may apply in one regime but fail in another depending on the
properties of the galaxies in one's sample.  Here, we have shown that
galaxies that lie around $L^{\ast}$ of the UV luminosity function have
dust obscuration -- as measured from {\em Spitzer}, {\em Herschel},
and VLA data -- that correlate with their UV slopes, and that this
correlation is similar to that observed for local starburst galaxies.

The stacked {\em Herschel} data do not directly indicate the intrinsic
dispersion in the relation between UV slope and dustiness.  However,
an indirect estimate of this scatter comes from an analysis of the
$24$\,$\mu$m data.  Specifically, 109 of 311 galaxies in the larger
(and multiple field) sample of \citet{reddy10a} are detected at
$24$\,$\mu$m.  Using a survival analysis to take into account both
detections and non-detections, \citet{reddy10a} found a dispersion of
$\approx 0.40$\,dex between dust attenuation, $\lir/\luv$, and rest-UV
slope, $\beta$.  If the scatter in the $\mirlum$-to-$\lir$ ratio is
similar to that found by \citet{elbaz11} for lower redshift ($z\la
2.0$) galaxies with $\lir\la 10^{12}$\,L$_{\odot}$ (having a
$1$\,$\sigma$ dispersion of $\approx 0.1$\,dex), then the implied
total dispersion in the relation between dustiness and UV slope is
$\approx 0.45$\,dex.  The correspondence between $\lir/\luv$ and
$\beta$ to within a factor of $\approx 3$ thus verifies the
applicability of the \citet{meurer99} and \citet{calzetti00}
attenuation curves for galaxies with $\lir \la 10^{12}$\,L$_{\odot}$
at $z\sim 2$.  The UV attenuation curve is sensitive primarily to the
geometry of dust and stars within galaxies, and/or variations in dust
composition.  Hence, the correspondence of the UV attenuation curves
between local starbursts and $z\sim 2$ $L^{\ast}$ galaxies implies a
remarkable similarily in the processes that give rise to the spatial
distribution of dust and stars and the dust composition in galaxies
over $\approx 10$\,billion years of cosmic history.

\subsection{Comparison with Recent Studies}

Using the {\em Herschel} data to directly probe the thermal dust
emission, we have shown that the local correlation between UV slope
and dust obscuration remains valid for typical star-forming galaxies
at $z\sim 2$.  In the following, we compare our results with several
recent studies of the dust attenuation of high redshift galaxies.

\subsubsection{Radio Emission from UV-selected Galaxies}

\citet{carilli08} stack the radio emission of $z\sim 3$ LBGs in the
COSMOS field and deduce that SFR$_{\rm radio}$/SFR$_{\rm UV} = 1.8\pm
0.4$, where SFR$_{\rm radio}$ is derived using the calibration of
\citet{yun01}.  This calibration is based on equating the local star
formation rate density with the integral of the radio luminosity
function, so in this case the radio SFR should represent a total SFR
(but see below), including obscured and unobscured components.  The
factor of $1.8$ from the \citet{carilli08} study is significantly
smaller than the factor of $\approx 5$ computed for our UV-selected
sample at $z\sim 2$.

\citet{carilli08} discuss several possibilities for the suppression of
radio flux with respect to SFR at $z\sim 3$.  For a consistent
comparison, we assess their results using the same calibration used in
this analysis.  In particular, we employ the \citet{bell03}
correlation between total infrared luminosity ($\lir$) and specific
luminosity at 1.4\,GHz.  Doing so, the stacked median radio luminosity
of LBGs in the COSMOS field, $L_{1.4} = 5.1\times
10^{29}$\,erg\,s$^{-1}$\,Hz$^{-1}$, corresponds to $\lir \approx
2.2\times 10^{11}$\,L$_{\odot}$.  These values are essentially
identical to those determined for our $L^{\ast}$ sample at $z\sim 2$:
$L_{1.4} = (5.2\pm1.0)\times 10^{29}$\,erg\,s$^{-1}$\,Hz$^{-1}$ and
$\lir = (2.2\pm0.3)\times 10^{11}$\,L$_{\odot}$
(Table~\ref{tab:lircomp}).  The {\em obscured} SFR corresponding to
the $\lir$ for the COSMOS LBGs, assuming the \citet{kennicutt98}
relation, is $\approx 38$\,M$_{\odot}$\,yr$^{-1}$.  Hence, the factor
that we compute to recover the total star formation rate from the UV
star formation for the \citet{carilli08} sample, assuming their value
of the unobscured SFR of $17$\,M$_{\odot}$\,yr$^{-1}$, is $1+38/17
\approx 3.2$.  This is close to a factor of two larger than the value
of 1.8 given in \citet{carilli08}.  This discrepancy results from the
fact that while the \citet{yun01} correlation between radio luminosity
and star formation rate gives a star formation rate that is in good
agreement with that inferred from $\lir$ (SFR\,$\simeq
38$\,M$_{\odot}$\,yr$^{-1}$), it underestimates the total SFR
(SFR$_{\rm IR}$+SFR$_{\rm UV}$) of $\simeq
55$\,M$_{\odot}$\,yr$^{-1}$.

\begin{figure}
\plotone{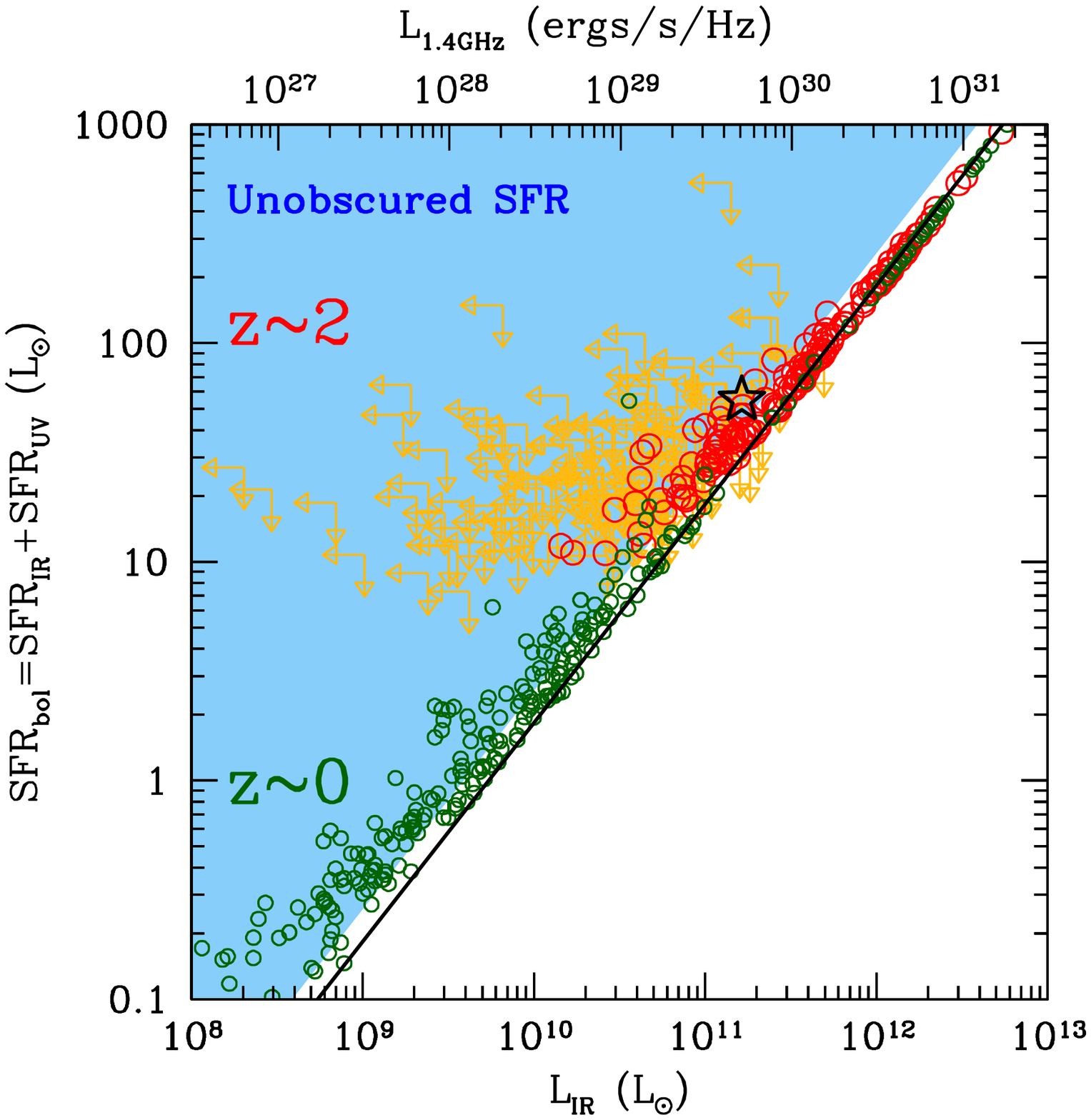}
\caption{ Total star formation rate versus infrared luminosity
  ($\lir$) for the sample of 392 UV-selected $z\sim 2$ galaxies of
  \citet{reddy10a} ({\em red circles and orange upper limits}), and
  the local samples of \citet{bell03} and \citet{huang09} ({\em dark
    green circles}).  The top axis shows the radio luminosity that
  corresponds to $\lir$ assuming the radio-IR correlation of
  \citet{yun01}, and the solid line denotes the relationship between
  radio luminosity and star formation rate derived in that study.  The
  shaded region covers the area where the unobscured star formation is
  at least $50\%$ of the obscured star formation.  The large open star
  denotes the position of the COSMOS LBGs from \citet{carilli08}.}
\label{fig:sfrir}
\end{figure}

To illustrate this point, we plot in Figure~\ref{fig:sfrir} the
relationship between {\em total} star formation rate and infrared
luminosity, adopting the \citet{kennicutt98} conversions between UV/IR
luminosity and star formation rate, for the sample of 392 $z\sim 2$
galaxies of \citet{reddy10a}, and for the local samples of
\citet{bell03} and \citet{huang09}.  The top axis in
Figure~\ref{fig:sfrir} indicates the radio luminosity that corresponds
to $\lir$ assuming the \citet{yun01} calibration, and the solid line
shows the relationship between radio luminosity and star formation
rate derived in that study.  The \citet{yun01} calibration is valid if
most of the star formation is obscured, as is the case for ULIRGs at
both $z\sim 2$ and $z\sim 0$.  However, a substantial fraction of the
LIRGs in the $z\sim 2$ sample have a significant contribution from
unobscured star formation, where the unobscured component is at least
$50\%$ of the obscured star formation.  The same is true for the
COSMOS LBG sample of \citet{carilli08}.  The ratio of obscured to
unobscured star formation rate is a strong function of total star
formation rate or bolometric luminosity (e.g., Figure~\ref{fig:sfrir};
\citealt{reddy10a}), and the UV component obviously cannot be
neglected for objects that have significant UV emission.
Figure~\ref{fig:sfrir} shows that one would significantly
underestimate the total star formation rate of LIRGs at $z\sim 2$
based on their IR emission alone (the \citealt{kennicutt98} relation
between SFR and $\lir$ is only valid in the optically thick limit), or
based on using the \citet{yun01} relationship between radio luminosity
and star formation rate.

Note also that \citet{carilli08} combine the {\em median} radio SFR
with the {\em mean} unobscured UV SFR to determine the effect of dust.
In general, mean luminosities will be larger than median ones for a
population drawn from a \citet{schechter76} luminosity function,
modulo sample incompleteness.  For our $L^{\ast}$ sample, the mean UV
luminosity is about $15\%$ larger than the median.  More importantly,
the {\em mean} unobscured UV luminosity of the \citet{carilli08}
sample of $17$\,M$_{\odot}$\,yr$^{-1}$ is about a factor of two larger
than the {\em median} UV SFR for our $L^{\ast}$ sample of $\approx
8.5$\,M$_{\odot}$\,yr$^{-1}$.  It is possible that the relatively
UV-bright LBGs of the \citet{carilli08} sample are somewhat less
attenuated than more typical (and less UV-luminous) LBGs at $z\sim
2-3$, though this is contrary to what has been found for UV-selected
galaxies at these redshifts \citep{reddy10a}.  Without further
analysis of their candidates, we conclude that the lower UV
obscuration factor deduced by \citet{carilli08} is likely due to the
brighter unobscured UV luminosity of their candidates, relative to the
obscured star formation.  What is clear from our spectroscopic sample
is that the stacked radio flux for UV-selected galaxies at $z\sim 2$
predicts an $\lir$ that is identical to that obtained using direct
measurements of thermal dust emission from the {\em Herschel} data,
and this value of $\lir$ implies a dust correction of a factor of
$\approx 5$.

\subsubsection{Investigations of the Extragalactic Background Light (EBL)}

A second and more recent study that has suggested obscuration factors
that are different from those predicted from the \citet{meurer99}
relation comes from an analysis of the infrared background based on
stacked {\em Spitzer} and BLAST data by \citet{chary10}.  These
authors estimate the extragalactic background light (EBL) contributed
from galaxies at $z\ga 1$ and conclude that the UV-based dust
corrections for typical star-forming galaxies (e.g., those selected by
their UV emission) must be lower than the \citet{meurer99} prediction
in order that their integrated emission does not violate the
background light constraints.  However, directly probing the thermal
dust emission of $L^{\ast}$ galaxies at $z\sim 2$, as we have done
here, shows conclusively that these galaxies have dust attenuations
that are similar, on average, to those computed based on the
\citet{meurer99} and \citet{calzetti00} attenuation curves
(Figure~\ref{fig:ebmv100}).  How can we reconcile these two results?

The EBL is sensitive to the average dust attenuation of {\em
  all} galaxies, not just the UV-bright ones studied here.  Indeed,
\citet{reddy09} and \citet{reddy10a} use physical arguments and
stacked {\em Spitzer} data to show that the dust obscuration of
UV-faint galaxies is lower than in UV-bright ones (see also next
section), and that this luminosity dependence implies that the average
dust obscuration, integrated over the entire luminosity function, can
be close to a factor of two lower than the mean dust obscuration found
for just the UV bright galaxies (see Table~5 of \citealt{reddy09}).
This does not necessarily imply a failure of the \citet{meurer99}
relation for typical star-forming galaxies at high redshift; it simply
means that the rest-frame UV slope also becomes bluer, on average, for
UV-faint galaxies, as the empirical evidence seems to indicate at
$z\sim 2.5$ \citep{bouwens09}.  What is clear from the present
analysis is that the direct measurements of the thermal dust emission
of UV-selected galaxies at $z\sim 2$ imply that the local UV
attenuation curve remains valid for these galaxies.

\subsubsection{Stacked {\em Herschel}/SPIRE Measurements for $z\sim 2$ UV-selected
Galaxies}

Strong source confusion in the {\em Herschel}/SPIRE 250, 350 and
500\,$\mu$m data make it very difficult to carry out stacking
experiments capable of reaching flux limits as faint as those we
expect for $L^\ast$ UV-selected galaxies at $z\sim2$ (approximately
1\,mJy -- see Figure~\ref{fig:avesed}).  After some experimentation,
we have chosen not to include those data in our analysis.
Nevertheless, \citet{rigopoulou10} stack SPIRE 250\,$\mu$m data for a
smaller sample of brighter, 24\,$\mu$m-detected, UV-selected galaxies
at $z\sim 2$.  These were taken from the sample of \citet{reddy06b},
also used here, but were limited to a 69 objects with individual
24\,$\mu$m detections.  That subsample is therefore likely to be more
IR-luminous, on average, than the larger sample considered here, where
individual detection at 24\,$\mu$m is not required.
\citet{rigopoulou10} measure a stacked 250\,$\mu$m flux of $f_{250} =
2.7\pm0.8$\,mJy, corresponding to a total infrared luminosity $\lir
\approx 4.2\times 10^{11}$\,L$_\odot$ at $\langle z \rangle \approx
2$.\footnote{We note that the stacked 250\,$\mu$m flux shown in
  Figure~2 of \citet{rigopoulou10} appears to be nearly 10$\times$
  fainter than the value cited in the text.  The authors state that
  they derive a total infrared luminosity $\lir = (1.5\pm 0.5) \times
  10^{11}$\,L$_\odot$ using the CE01 dust templates, but we are unable
  to reproduce this, finding instead a luminosity that is 2.8$\times$
  larger, based on the value $f_{250} = 2.7$\,mJy cited in the text.}
This value is similar to what we predict at 250\,$\mu$m when we stack
the 100 and 160\,$\mu$m data for the same set of galaxies (i.e., the
brighter objects with individual 24\,$\mu$m detections).  Given that
we have (1) controlled for issues of clustering and confusion at
$100$\,$\mu$m, and cross-checked the $160$\,$\mu$m fluxes with those
obtained at $100$\,$\mu$m, and (2) performed detailed simulations to
demonstrate the robustness of the stacked fluxes
(Section~\ref{sec:stacking}), we believe the PACS constraints on the
median IR SED to be robust.  With the PACS data we are able to
constrain the median IR SED for $L^\ast$ galaxies with lower average
infrared luminosities, irrespective of whether they are individually
detected at 24\,$\mu$m.  Future observations with the Atacama Large
Millimeter Array (ALMA) should provide higher resolution and more
sensitive observations of the Rayleigh-Jeans emission from typical
star-forming galaxies at high redshift.

\begin{figure*}
\plottwo{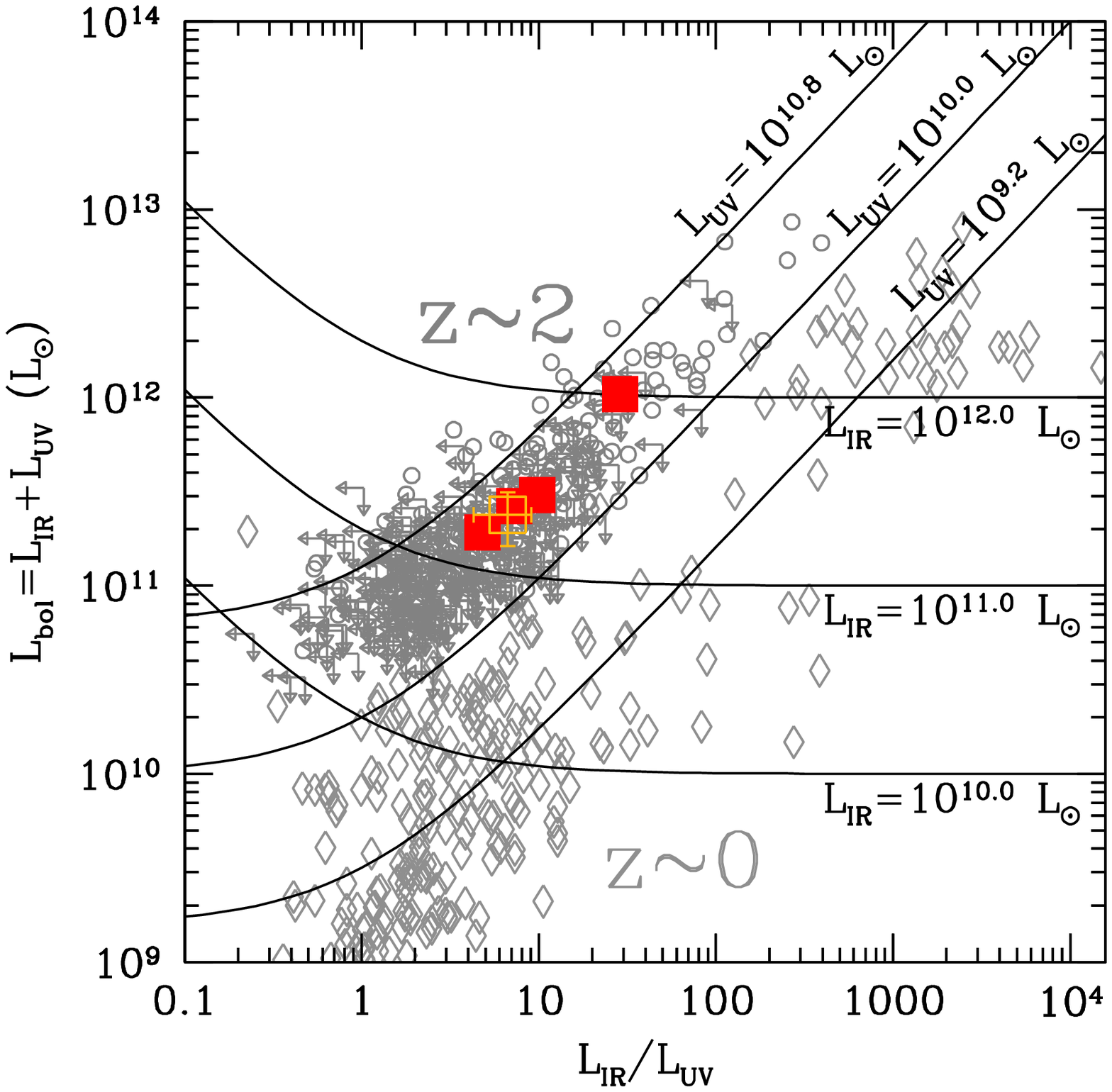}{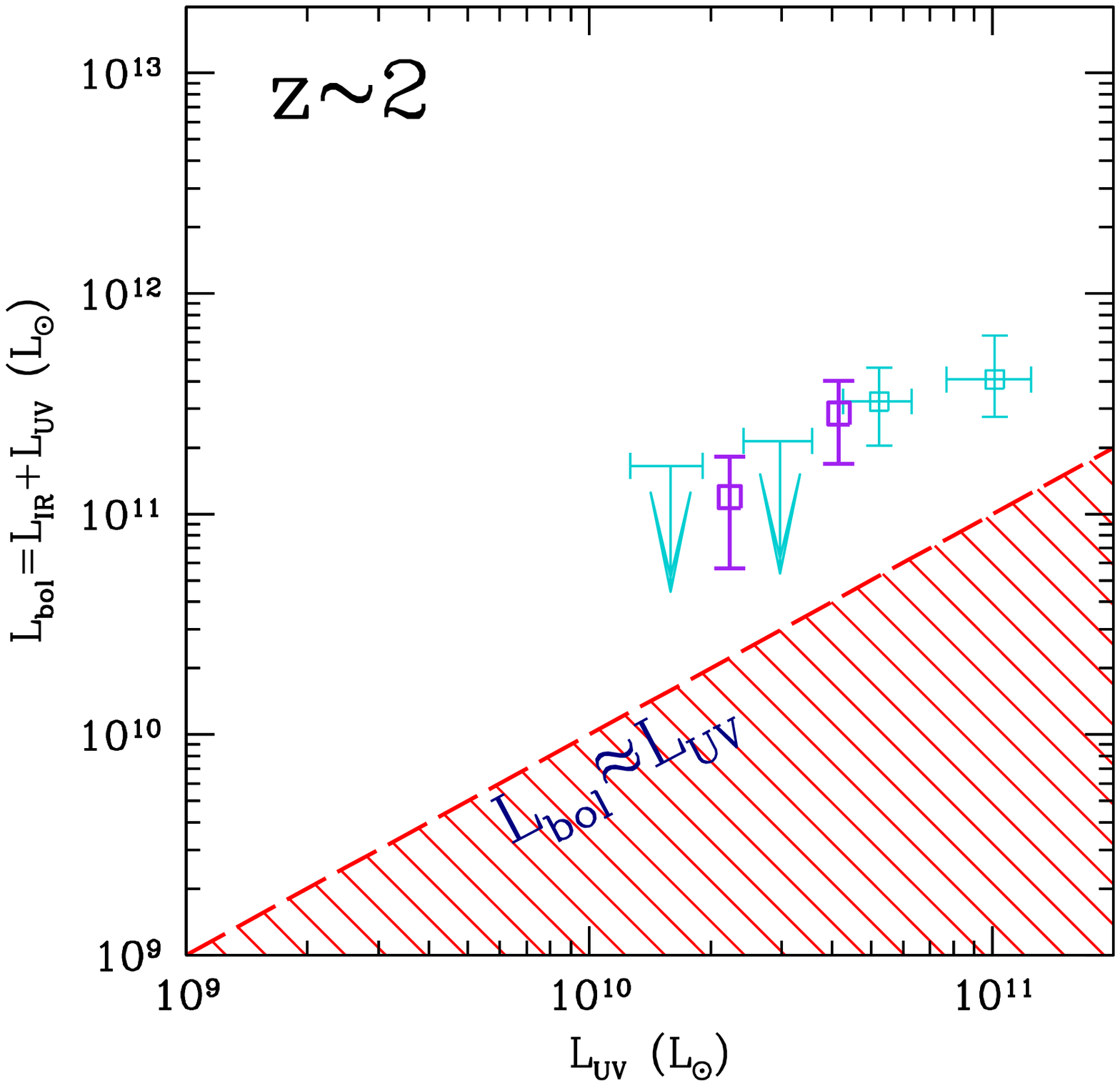}
\caption{Left: Bolometric luminosity, $\lbol \equiv \lir + \luv$, as a
  function of $\lir/\luv$ for a sample of local galaxies from
  \citet{bell03} and \citet{huang09}, shown by the open diamonds.  The
  sample of 392 UV-selected galaxies at $1.5\le z<2.6$ analyzed in
  \citet{reddy10a} are shown by the open circles and upper limits.
  The relationship between $\lbol$ and $\lir/\luv$ for Samples A, C,
  D, and E, based on the stacked {\em Herschel}, {\em Spitzer}, and
  VLA data, are denoted by the large filled squares (error bars are
  not shown for clarity).  The same quantity for the $L^{\ast}$ sample
  at $z\sim 2$ (Sample B) is shown by the large open square.  Right:
  Correlation between $\lbol$ and $\luv$, where the cyan points denote
  the results using $24$\,$\mu$m data and the \citet{reddy10a}
  calibration between $\mirlum$ and $\lir$.  The purple points
  indicate results from incorporating the {\em Spitzer}/MIPS
  $24$\,$\mu$m, {\em Herschel}/PACS 100 and 160\,$\mu$m, and VLA
  1.4GHz data to determine the median IR SED as a function of UV
  luminosity.}
\label{fig:bol}
\end{figure*}

\subsection{Dependence of Dust Attenuation on Stellar Population Age and 
Luminosity}

\subsubsection{Galaxies with ``Young'' ($\la 100$\,Myr) Stellar Population Ages}

Roughly $14\%$ of the galaxies in our entire sample are identified as
having a ``young'' stellar population with inferred ages of $\la
100$\,Myr.  Stacking these galaxies results in a formal nondetection
(S/N\,$\la 3$) in the {\em Herschel} and VLA stacks
(Table~\ref{tab:fluxes}).  The implied $3$\,$\sigma$ upper limit to
the dust obscuration suggests that these galaxies follow an
attenuation curve that is ``steeper'' than the typically assumed
\citet{meurer99} and \citet{calzetti00} attenuation curves
(Figure~\ref{fig:ebmv100}), in the sense that they exhibit a redder UV
slope for a given obscuration.  More stringent constraints on the dust
attenuation of these galaxies comes from their stacked $24$\,$\mu$m
emission.  Assuming the DH02 template that best matches the median
$24$\,$\mu$m flux of the ``young'' galaxies implies $\lir/\luv \approx
1.55\pm 0.35$.  The median UV slope of these galaxies of $\beta =
-1.30$ then implies that their dust obscurations are consistent with
the SMC attenuation curve.  The deviation of such young, high-redshift
galaxies from the standard attenuation curve is a result first noted
in \citet{reddy06a} and further investigated in \citet{reddy10a}.  IRS
spectrocopy of a couple of young lensed LBGs provide additional
evidence that such young galaxies are less dusty at a given UV slope
\citep{siana08, siana09} than their older and more typical
counterparts at the same redshifts \citep{reddy10a}.

The small carbonaceous dust grains giving rise to the mid-IR emission
are believed to be produced in AGB stars over a longer timescale than
the Type II SNe responsible for the large grains that emit thermally
in the infrared (e.g., \citealt{galliano08}).  It is therefore
relevant to ask whether evolution in the small-to-large dust grain
ratio may be responsible for the observed differences in attenuation
curve with stellar population age.  Based on an X-ray stacking
analysis and an examination of the theoretical PAH metallicity from
dust models, \citet{reddy10a} argued that the small-to-large dust
grain ratio is unlikely to change significantly over the relatively
narrow range in metallicity probed by the UV-selected sample.  IRS
spectroscopy of at least a couple of young lensed LBGs at
high-redshift also indicates that their ratios of mid-to-far IR
emission are similar to those found for local starburst galaxies
\citep{siana08, siana09} and similar to the ratios observed for $z\sim
2$ galaxies \citep{reddy06a, reddy10a}.  The {\em Herschel}
observations, which probe the thermal dust emission at $z\sim 2$, are
consistent with these findings based on $24$\,$\mu$m emission alone.

If the difference in attenuation curve is due purely to geometrical
effects, then it suggests that the dust covering fraction of
high-redshift galaxies evolves significantly with time.  A possible
simple scenario is one in which the first generation of stars quickly
pollutes the ISM with dust and metals over a dynamical timescale of a
few tens of millions of years, at which point much of the dust is
foreground to the stars and the dust covering fraction is high.  As
star formation proceeds and either becomes spatially extended or is
able to drive outflows sufficient in momentum to perturb the dust/gas
distribution of the ISM, the resulting attenuation becomes patchier.
So, we might expect in this situation that galaxies would gradually
transition from a steep attenuation curve like that of the SMC to a
grayer starburst attenuation curve (e.g., \citealt{meurer99,
  calzetti00}; see also discussion in \citealt{buat11}).  Detailed
studies of the variation of the interstellar absorption lines (as a
proxy for the dust covering fraction) with dust attenuation will be
needed to test this scenario.

\subsubsection{Variation in Dust Attenuation with Bolometric and UV Luminosity}

MIPS $24$\,$\mu$m studies have shown that $z\sim 2$ galaxies follow a
tight trend between bolometric luminosity and dust attenuation (e.g.,
\citealt{reddy10a} and references therein), similar to that observed
locally and at $z\sim 1$ \citep{burgarella09, buat07, buat09, wang96}.
However, the normalization of this trend depends on redshift, such
that at a given bolometric luminosity, galaxies at high redshift are
on average less dusty than local ones \citep{reddy06a, reddy08,
  reddy10a}.  This result is not unique to our UV-selected sample.
Rest-frame optical, near-IR, and submillimeter selected galaxies at
$z\sim 2$ also show a similar offset in obscuration per unit star
formation compared with local galaxies \citep{reddy06a}.
\citet{reddy10a} demonstrate that this relationship is likely driven
by metallicity evolution.  The stacked {\em Herschel} and radio data
confirm these previous results (Figure~\ref{fig:bol}).  This is not
surprising given the similarity in $\lir$ and dust attenuation
inferred in this study with those inferred for $L^{\ast}$ galaxies
based on $24$\,$\mu$m data alone.  The combined {\em Herschel}, {\em
  Spitzer}, and VLA data confirm that $L^{\ast}_{\rm UV}$ galaxies at
$z\sim 2$ are a factor of $\approx 10$ less dusty than galaxies with
similar bolometric luminosities in the local universe.  Similarly,
$L^{\ast}_{\rm UV}$ galaxies at $z\sim 2$ are a factor of $\approx
20-30$ times more bolometrically-luminous than local galaxies with a
similar dust obscuration (Figure~\ref{fig:bol}).  These evolutionary
effects can also be seen in Figure~\ref{fig:sfrir}.  In the local
universe, it is only for galaxies with $\lir\la 10^{11}$\,L$_{\odot}$
(i.e., galaxies fainter than LIRGs) where the unobscured star
formation begins to contribute significantly to the total star
formation rate.  In contrast, most lower luminosity LIRGs at $z\sim 2$
have a significant fraction of unobscured star formation, as indicated
by the plume of $z\sim 2$ galaxies extending away from the SFR$_{\rm
  bol}=$SFR$_{\rm IR}$ line in Figure~\ref{fig:sfrir}, indicating that
LIRGs at $z\sim 2$ are more UV-transparent relative to LIRGs in the
local Universe.

Finally, we note that \citet{reddy10a} use a $24$\,$\mu$m analysis of
a larger sample of UV-selected galaxies at $z\sim 2$ to show that the
fraction of $24$\,$\mu$m detections of these galaxies decreases by a
factor of two proceeding from UV-bright ($\luv \approx
10^{11}$\,L$_\odot$) to UV-faint ($\luv \approx 10^{10}$\,L$_\odot$)
galaxies (see Figure~14 in \citealt{reddy10a}).  This UV luminosity
trend in $24$\,$\mu$m detection fraction implies that UV-faint
galaxies are on average less infrared-luminous than their UV-bright
counterparts, a result that is further confirmed by stacking the
$24$\,$\mu$m emission of galaxies in bins of UV luminosity.  A
stacking of the combined {\em Spitzer}, {\em Herschel} and VLA data as
a function of UV luminosity shows a similar trend, indicating that
UV-faint galaxies are also less bolometrically luminous than UV-bright
galaxies (Figure~\ref{fig:bol}).  The actual dust attenuation,
$\lir/\luv$, is roughly constant (within the uncertainties) with
decreasing UV luminosity, as both UV and IR luminosities decrease in
tandem.  In any case, it is clear from these observations that care
must be taken when inferring the average dust obscuration and $\lir$,
and their effect on global quantities like the star formation rate
density, given the UV luminosity dependence of these quantities
\citep{reddy08, reddy09}.

\section{Conclusions}

We have used the deep $100$ and $160$\,$\mu$m data of the GOODS-{\em
  Herschel} Open Time Key Program, supplemented with deep {\em
  Spitzer}/MIPS $24$\,$\mu$m and VLA 1.4\,GHz radio imaging, to
investigate the infrared luminosities and dust obscuration of typical
star-forming galaxies at high redshift.  We focus on the median
stacked mid-infrared, far-infrared, and radio fluxes of a sample of
$146$ UV-selected galaxies with spectroscopic redshifts $1.5\le z_{\rm
  spec}<2.6$ in the GOODS-North field.  These galaxies have
luminosities around $L^{\ast}$ of the UV luminosity function at these
redshifts.

Because these galaxies are individually undetected at $100$,
$160$\,$\mu$m, and $1.4$\,GHz, we perform median stacking analyses to
measure their average fluxes.  Three tests are performed to verify the
robustness of our stacked results.  The first test measured the
effects of clustering. The second test determined the chance
probability of measuring a stacked flux as high as the one obtained
for the target galaxies.  The third test allows us to measure the
biases and uncertainties in stacked flux measurements by stacking on
the positions of artificial sources added to the images.

To interprete these fluxes, we consider a variety of dust templates,
including those of \citet{elbaz11}, \citet{chary01}, \citet{dale02},
and \citet{rieke09}, as well as the $24$\,$\mu$m-$\lir$ and
radio-infrared correlations of \citet{reddy10a} and \citet{bell03},
respectively.  Fitting these templates to the bias-corrected fluxes
and infrared colors reveals that $L^{\ast}_{\rm UV}$ galaxies at
$z\sim 2$ with UV luminosities $\luv \ga 10^{10}$\,L$_{\odot}$ have a
median infrared luminosity of $\lir= (2.2\pm 0.3)\times
10^{11}$\,L$_{\odot}$.  Galaxies in our sample exhibit $\lir/\mirlum$
(IR8) ratios that are a factor of $\approx 2$ larger than those found
for most star-forming galaxies at lower redshifts, likely due to the
fact these UV-selected galaxies are relatively compact for their
infrared luminosity.  One possibility is that the roughly constant IR8
ratio observed for most (main sequence) galaxies with redshifts $z\la
2.0$ likely shifts towards larger values at $z\ga 2.0$, due to the
fact that the galaxies are on average smaller for their IR luminosity
relative to lower redshift galaxies.

Stacking the {\em Spitzer}, {\em Herschel}, and VLA data as a
function of UV spectral slope, $\beta$, and bolometric luminosity,
$\lbol$, indicates that galaxies with redder $\beta$ and higher
$\lbol$ have on average larger infrared luminosities, in accord with
expectations based on trends found for local star-forming galaxies.
Based on the $\lir$, we proceed to examine the dust attenuation,
$\lir/\luv$, and the dust correction factor (i.e., the factor required
to recover the bolometric SFR from the unobscured UV SFR), 1+
SFR$_{\rm IR}$/SFR$_{\rm UV}$, for galaxies in our sample.  For
$L^{\ast}$ galaxies at $z\sim 2$, we find a dust attenuation of
$\lir/\luv = 7.1\pm1.1$, which corresponds to a dust correction factor
of $5.2\pm 0.6$, implying that $\simeq 80\%$ of the star formation is
obscured.  This result is consistent with those found in previous
studies of the dust-corrected UV and H$\alpha$, $24$\,$\mu$m, and radio
and X-ray stacks of the same galaxies examined here \citep{reddy04,
  reddy06a, reddy10a}.

We have examined the relationship between the UV spectral slope,
$\beta$, and dustiness for $z\sim 2$ galaxies.  A small fraction
($\approx 14\%$) of galaxies are identified as having stellar
population ages $\la 100$\,Myr from fitting stellar population SEDs to
the observed optical through near-IR photometry.  The upper limit in
dust attenuation for these ``young'' galaxies suggests that they may
follow a ``steeper'' attenuation curve than the one observed for local
starburst galaxies \citep{meurer99, calzetti00}, in the sense that
they are less dusty (i.e., have lower $\lir/\luv$ ratios) for a given
$\beta$ than older and more typical galaxies at the same redshift.
This result was first noted in the $24$\,$\mu$m analyses of
\citet{reddy06a} and further investigated in \citet{reddy10a}, and may
be attributable to the higher dust covering fractions in young
galaxies.  When considering more typical galaxies with ages $\ga
100$\,Myr, we find that their median $\lir/\luv$ increases as $\beta$
becomes redder, and that this correlation is essentially identical to
that found for local starbursts \citep{meurer99, calzetti00}.
Comparison of the {\em Herschel} results at $z\sim 2$ with
measurements of local galaxies confirms the previously found trends
that imply that LIRGs at $z\sim 2$ are more UV transparent (i.e., less
dusty) than LIRGs in the local universe, and that UV-faint galaxies at
$z\sim 2$ have lower $\lir$ and hence fainter bolometric luminosities
than UV-bright galaxies at $z\sim 2$ (e.g., see also
\citealt{reddy10a}).  Based on the direct {\em Herschel} measurements
of the rest-frame $\simeq 30$ and $50$\,$\mu$m thermal dust emission,
we find that the local UV attenuation curve holds at $z\sim 2$ for
galaxies with $\lbol \la 10^{12}$\,L$_{\odot}$, suggesting a
remarkable similarity in the processes governing star formation and
dust production over the last $10$\,billion years of cosmic time.

We have made significant progress in evaluating the effects of dust in
typical star-forming galaxies at $z\sim 2$ as traced by thermal
infrared emission.  However, even the superb sensitivity and
resolution of {\em Herschel} are not sufficient to individually detect
$L^{\ast}$ galaxies at these redshifts.  The most accurate estimate of
dust luminosity, and hence total star formation rate, can only come
from combining {\em individual} measurements of UV and IR
luminosities.  Though the observations will be targeted, ALMA promises
to extend our knowledge of the dust continuum luminosities of {\em
  individual} star-forming galaxies at $z\sim 2$, thus providing the
key ingredient to combine with UV measurements.  These developments
point to a wealth of forthcoming information on the properties of dust
in typical star-forming galaxies at high redshift.  Ultimately, a
robust characterization of the relationship between infrared and short
wavelength (UV, H$\alpha$) emission will be crucial for inferring dust
attenuation and total star formation rates of the very faint and/or
very high redshift galaxies that will be inaccessible to even the next
generation of long wavelength observatories.

\acknowledgements

Support for NAR was provided by NASA through Hubble Fellowship grant
HST-HF-01223.01 awarded by the Space Telescope Science Institute,
which is operated by the Association of Universities for Research in
Astronomy, Inc., for NASA, under contract NAS 5-26555.  This work is
based on observations made with the {\em Herschel Space Observatory},
a European Space Agency Cornerstone Mission with significant
participation by NASA.  Support for this work was provided by NASA
through an award issued by JPL/Caltech.


\end{document}